\documentclass[12pt]{article}
\usepackage{epsf}
\usepackage{amsmath}
\usepackage{graphics}

\renewcommand{\thefootnote}{\#\arabic{footnote}}
\newcommand{\gtrsim}{ \mathop{}_{\textstyle \sim}^{\textstyle >} }
\newcommand{\lesssim}{ \mathop{}_{\textstyle \sim}^{\textstyle <} }

\begin{document}

\setcounter{footnote}{0}
\begin{titlepage}

\begin{center}

\hfill hep-ph/0507245\\
\hfill OU-TAP-261\\
\hfill TU-749\\
\hfill July, 2005\\

\vskip .5in

{\Large \bf
Big-Bang Nucleosynthesis with Unstable Gravitino and
Upper Bound on the Reheating Temperature
}

\vskip .45in

{\large
Kazunori Kohri$^{a,b}$,
Takeo Moroi$^c$ and Akira Yotsuyanagi$^c$
}

\vskip .45in

{\em
$^a$Harvard-Smithsonian Center for Astrophysics, 60 Garden Street\\
Cambridge, MA 02138, U.S.A.
}

\vskip .2in

{\em
$^b$Department of Earth and Space Science, Graduate School of
Science\\ 
Osaka University, Toyonaka  560-0043, Japan
}

\vskip .2in

{\em
$^c$Department of Physics, Tohoku University\\
Sendai 980-8578, Japan \\
}

\end{center}

\vskip .4in

\begin{abstract}

We study the effects of the unstable gravitino on the big-bang
nucleosynthesis.  If the gravitino mass is smaller than $\sim 10~{\rm
TeV}$, primordial gravitinos produced after the inflation are likely
to decay after the big-bang nucleosynthesis starts, and the light
element abundances may be significantly affected by the hadro- and
photo-dissociation processes as well as by the $p\leftrightarrow n$
conversion process.  We calculate the light element abundances and
derived upper bound on the reheating temperature after the inflation.
In our analysis, we calculate the decay parameters of the gravitino
(i.e., lifetime and branching ratios) in detail.  In addition, we
performed a systematic study of the hadron spectrum produced by the
gravitino decay, taking account of all the hadrons produced by the
decay products of the gravitino (including the daughter
superparticles).  We discuss the model-dependence of the upper bound
on the reheating temperature.

\end{abstract}

\end{titlepage}

\renewcommand{\thepage}{\arabic{page}}
\setcounter{page}{1}
\renewcommand{\thefootnote}{\#\arabic{footnote}}
\renewcommand{\theequation}{\thesection.\arabic{equation}}

\section{Introduction}
\setcounter{equation}{0}

Low-energy supersymmetry, which is one of the most prominent
candidates of the physics beyond the standard model, may significantly
affect the evolution of the universe.  One reason is that, assuming
$R$-parity conservation, the lightest superparticle (LSP) is stable,
which becomes a well-motivated candidate of the cold dark matter.
Another famous reason, which is very important in the framework of
local supersymmetry (i.e., supergravity), is that there exist various
possible very weakly interacting particles in the particle content.
The most important example is the gravitino, which is the gauge field
for the local supersymmetry.

Since the gravitino is the superpartner of the graviton, its
interaction is suppressed by inverse powers of the gravitational scale
and hence its interaction is very weak.  Even though the gravitino is
very weakly interacting, however, it can be produced by the scattering
processes of the standard-model particles (and their superpartners) in
the thermal bath in the early universe.  Once produced, the gravitino
decays with very long lifetime.  In particular, if the gravitino mass
is smaller than $\sim 20\ {\rm TeV}$, the lifetime becomes longer than
$\sim 1\ {\rm sec}$ and hence the primordial gravitinos decay after
the big-bang nucleosynthesis (BBN) starts.  The (unstable) gravitino
is expected to be relatively heavy, and its decay releases energetic
particles which cause dissociation processes of the light elements
generated by the standard BBN reactions.  Since the standard BBN
scenario more or less predicts light-element abundances consistent
with the observations, those dissociation processes are harmful and,
if the abundance of the primordial gravitinos are too much, resultant
abundances of the light elements become inconsistent with the
observations.  Such argument provides significant constraints on the
primordial abundance of the gravitino \cite{Weinberg:zq}.\footnote
{Constraints on the case with the gravitino LSP have been considered in
Refs.\ \cite{Moroi:1993mb,sWIMP}.}

In the inflationary scenario, which is also strongly suggested by the
WMAP data \cite{WMAP}, gravitinos are once diluted by the entropy
production after the inflation.  Even in this case, however,
gravitinos are produced again by the scattering processes of the
particles in the thermal bath.  Since the total amount of the
gravitinos produced by the scattering processes is approximately
proportional to the reheating temperature $T_{\rm R}$, the BBN
scenario is too much affected to be consistent with the observations
if the reheating temperature is high.  In the past, the BBN
constraints on the unstable gravitino have been intensively studied
\cite{GravClassic,KawMor,Kawasaki:1994bs,Protheroe:dt,Holtmann:1998gd,
Jedamzik:1999di,Kawasaki:2000qr,Kohri:2001jx,Cyburt:2002uv,
Jedamzik:2004er,Kawasaki:2004yh,Kawasaki:2004qu,
Ellis:2005ii}.\footnote{
In order to relax the constraints, several scenarios have been
studied.  In Ref.~\cite{Okada:2004mh}, it was discussed that the
modification of the expansion rate by the 5D effect in the brane-world
cosmology, which may reduce the abundance of the gravitino. In
Ref.~\cite{KYY}, they studied possibilities of the dilution of the
gravitino by the late-time entropy production due to the decaying
moduli without newly producing many gravitinos.}

Recently, Kawasaki and two of the present authors (K.K.  and T.M.)
have studied the constraints on the unstable long-lived particle from
the BBN in detail \cite{Kawasaki:2004yh,Kawasaki:2004qu}; in
particular, in this paper, effects of the hadrons produced by the
decay of such unstable particles (as well as the effects of the
photo-dissociation) were systematically studied,\footnote
{For old studies on the effects of the hadronic decay modes, see also
\cite{Reno:1987qw,DimEsmHalSta}.}
and general constraints on the primordial abundance of such unstable
particles were presented.  Then, the results were applied to the case
of the unstable gravitino and the upper bound on the reheating
temperature was obtained for several simple cases.

In \cite{Kawasaki:2004yh,Kawasaki:2004qu}, however, several
simplifications and assumptions are made for the properties of the
gravitino.  First of all, several very simple decay patterns of the
gravitino were considered, which are applicable for very specific mass
spectrum of the superparticles.  In addition, for the hadronic
branching ratio of the gravitino, only several typical values were
used to obtain the constraint.  Furthermore, for some cases (in
particular, for the case where the gravitino dominantly decays into
the gluon and the gluino), effects of the hadrons emitted from the
superparticles (like the gluino) were neglected.  Thus, it is
desirable to perform more detailed and complete analysis of the upper
bound on the reheating temperature with the unstable gravitino.

In this paper, we study the effects of the unstable gravitinos on the
BBN, paying particular attention to the properties of the gravitino.
Compared to the previous works, decay processes of the gravitino (and
the decay chains of the decay products including the superparticles)
are precisely and systematically studied.  As a result, energy spectra
of the hadrons (in particular, proton, neutron, and pions) produced by
the gravitino decay are studied in detail for various mass spectrum of
the superparticles.

The organization of this paper is as follows.  In Section
\ref{sec:model}, we present the model we consider and summarize the
important parameters for our analysis.  In Section \ref{sec:decay},
detail of the decay processes of the gravitino is discussed.  Some of
the important issues in our analysis, which is the secondary decays of
the daughter particles from the gravitino decay and their
hadronization processes, are discussed in Section
\ref{sec:hadronization}.  Then, in Section \ref{sec:abundance},
outline of our calculation of the light-element abundances is
discussed.  Our main results are given in Section \ref{sec:results},
and Section \ref{sec:conclusions} is devoted for conclusions and
discussion.

\section{Model}
\label{sec:model}
\setcounter{equation}{0}

In this paper, we adopt the minimal particle content to derive the
upper bound on the reheating temperature.  Thus, the model we consider
includes the particles in the minimal supersymmetric standard model
(MSSM) as well as the gravitino.  These particles are listed in Table
\ref{table:plist}.

\begin{table}[t]
  \begin{center}
    \begin{tabular}{ll}
      \hline\hline
      Particles & Notation \\
      \hline
      Gravitino & $\psi_\mu$ \\
      Neutralinos & $\chi_1^0$, $\chi_2^0$, $\chi_3^0$, $\chi_4^0$ \\
      Charginos   & $\chi_1^\pm$, $\chi_2^\pm$ \\
      Gluino      & $\tilde{g}$\\
      Squarks     & $\tilde{q}$ $=$ $\tilde{u}_{\rm L}$, $\tilde{u}_{\rm R}$,
      $\tilde{d}_{\rm L}$, $\tilde{d}_{\rm R}$,
      $\tilde{s}_{\rm L}$, $\tilde{s}_{\rm R}$,
      $\tilde{c}_{\rm L}$, $\tilde{c}_{\rm R}$,
      $\tilde{b}_1$, $\tilde{b}_2$, $\tilde{t}_1$, $\tilde{t}_2$\\
      Sleptons    & $\tilde{e}_{\rm L}$, $\tilde{e}_{\rm R}$,
      $\tilde{\mu}_{\rm L}$, $\tilde{\mu}_{\rm R}$,
      $\tilde{\tau}_1$, $\tilde{\tau}_2$, 
      $\tilde{\nu}_{e_{\rm L}}$, 
      $\tilde{\nu}_{\mu_{\rm L}}$, $\tilde{\nu}_{\tau_{\rm L}}$\\
      Photon      & $\gamma$ \\
      Weak bosons & $Z$, $W^\pm$\\
      Gluon       & $g$\\
      Neutral CP even Higgses & $h$, $H$\\
      CP odd Higgs & $A$\\
      Charged Higgs & $H^\pm$\\
      Quarks      & $q$ $=$ $u$, $d$, $s$, $c$, $b$, $t$\\
      Leptons     & $e$, $\mu$, $\tau$, $\nu_e$, $\nu_\mu$, $\nu_\tau$\\
      \hline\hline
    \end{tabular}
    \caption{List of particles in the mass-eigenstate bases.}
    \label{table:plist}
  \end{center}
\end{table}

In order to precisely calculate the decay rate of the gravitino, it is
necessary to obtain the mass eigenvalues and mixing parameters of the
superparticles.  Thus, we briefly summarize the relation between the
gauge-eigenstate and mass-eigenstate bases here.  

We start with the neutralino sector.  With the $SU(2)_L$ and $U(1)_Y$
gaugino masses $M_2$ and $M_1$ as well as the supersymmetric Higgs
mass $\mu_H$, the mass matrix of the neutralinos is given in the
form\footnote
{We used the convention of \cite{Haber:1984rc}.}
\begin{eqnarray}
    {\cal M}_{\chi^0} = \left( 
        \begin{array}{cccc}
            M_1 & 0  
            & -m_Z\sin\theta_W\cos\beta &  m_Z\sin\theta_W\sin\beta \\
            0   &M_2 
            &  m_Z\cos\theta_W\cos\beta & -m_Z\cos\theta_W\sin\beta \\
            -m_Z\sin\theta_W\cos\beta & m_Z\cos\theta_W\cos\beta 
            & 0   &-\mu_H \\
            m_Z\sin\theta_W\sin\beta &-m_Z\cos\theta_W\sin\beta 
            & -\mu_H& 0   \\
        \end{array} \right),
    \nonumber \\
\end{eqnarray}
where $m_Z$ is the $Z$-boson mass while $\theta_W$ is the weak mixing
angle, and $\tan\beta$ is the ratio of the vacuum expectation value of
up- and down-type Higgs bosons.  This mass matrix is diagonalized by
a unitary matrix $U_{\chi^0}$ as
\begin{eqnarray}
U_{\chi^0}^* {\cal M}_{\chi^0} U_{\chi^0}^{-1} = 
{\rm diag}
(m_{\chi_1^0}, m_{\chi_2^0}, m_{\chi_3^0}, m_{\chi_4^0}).
\end{eqnarray}
In addition, for the chargino sector, the mass matrix is given by
\begin{eqnarray}
    {\cal M}_{\chi^\pm} = \left( 
        \begin{array}{cc}
            M_2 & \sqrt{2}m_W\cos\beta \\
            \sqrt{2}m_W\sin\beta & \mu_H 
        \end{array} \right),
\end{eqnarray}
where $m_W$ is the $W^\pm$-boson mass.  We define the unitary matrices
diagonalizing ${\cal M}_{\chi^\pm}$ as $U_{\chi^+}$ and $U_{\chi^-}$:
\begin{eqnarray}
U_{\chi^+}^* {\cal M}_{\chi^\pm} U_{\chi^-}^{-1} = 
{\rm diag}
(m_{\chi_1^\pm}, m_{\chi_2^\pm}).
\end{eqnarray}

For the neutral Higgs boson, the gauge-eigenstates (i.e., the up-type
Higgs $H_u^0$ and down-type Higgs $H_d^0$) and the mass-eigenstates
are related by using the mixing angle $\alpha$:
\begin{eqnarray}
    \left( \begin{array}{c}
            H \\ h
        \end{array} \right) =
    \sqrt{2}
    \left( \begin{array}{cc}
            \cos\alpha & \sin\alpha \\
            - \sin\alpha & \cos\alpha
        \end{array} \right)
    \left( \begin{array}{c}
            {\rm Re} (H_d^0) - v_1 \\
            {\rm Re} (H_u^0) - v_2
        \end{array} \right),
    \label{defalpha_TM}
\end{eqnarray}
where $v_1$ and $v_2$ are vacuum expectation values of  $H_d^0$ and
$H_u^0$, respectively.

In addition, we have to consider the mixings in the squark and slepton
mass matrices.  For simplicity, we do not consider the generation
mixing in the squark and slepton sector.  We also neglect the
left-right mixing in the first and second generation squarks and
sleptons since such mixing is small in many class of models, in
particular, in the minimal supergravity (mSUGRA) type models which we
adopt in our analysis.  For the third-generation squarks and sleptons,
we take account of the effects of the left-right mixing.  Such mixing
is parameterized by unitary matrices $U_{\tilde{t}}$, $U_{\tilde{b}}$,
and $U_{\tilde{\tau}}$ which diagonalize the mass-squared matrices of
the squarks and sleptons:
\begin{eqnarray}
U_{\tilde{f}} {\cal M}^2_{\tilde{f}} U_{\tilde{f}}^{-1} =
{\rm diag}
(m_{\tilde{f}_1}^2, m_{\tilde{f}_2}^2) ~~~:~~~
\tilde{f} = \tilde{t}, \tilde{b}, \tilde{\tau}.
\end{eqnarray}

In our analysis, we consider the case where the gravitino is unstable
and the LSP is contained in the MSSM particles.  Since the charged or
colored LSP is disfavored, we consider the case where the LSP is the
lightest neutralino $\chi_1^0$.  Consequently, the gravitino is
assumed to be heavier than $\chi_1^0$.  Of course, the gravitino may
be heavier than other superparticles and hence the lifetime and
branching ratios for possible decay modes of the gravitino depend on
the mass spectrum.  We calculate these quantities in detail, as we
explain below.

In order to calculate the decay rate of the gravitino, it is necessary
to fix the mass spectrum and the mixing matrices in the MSSM sector.
Although the effects of the gravitino on the BBN can be calculated for
arbitrary mass spectrum of the MSSM particles, it is not practical to
study all the possible cases since there are very large number of
parameters in the MSSM sector.  Thus, we adopt a simple
parameterization of the SUSY breaking parameters, that is, the
mSUGRA-type parameterization of the soft SUSY breaking parameters.  We
(parameterize the MSSM parameters by using unified gaugino mass
$m_{1/2}$, universal scalar mass $m_0$, universal coefficient for the
tri-linear scalar coupling $A_0$, ratio of the vacuum expectation
values of two Higgs bosons $\tan\beta$, and supersymmetric Higgs mass
$\mu_H$.  Then, the properties of the superparticles (including the
gravitino) are determined once these parameters as well as the
gravitino mass $m_{3/2}$ are fixed and, consequently, we can derive
the upper bound on the reheating temperature.  Notice that, although
we adopt the simple parameterization of the soft SUSY breaking
parameters, our analysis is applicable to more general cases as far as
the gravitino is heavier than one of the MSSM superparticles (like the
lightest neutralino).

Even with the mSUGRA parameterization of the MSSM parameters, the
whole parameter space is still too large to be completely studied.
Thus, in this paper, we pick up several typical mSUGRA points and
derive the constraints for these points.  In particular, we pick up
points where the thermal relic density of the LSP becomes consistent
with the dark matter density determined by the WMAP observation
\cite{WMAP}.  The points we consider are listed in Table
\ref{table:mSUGRA}.  For all the cases, we checked that the lightest
neutralino becomes the LSP (if the gravitino mass is larger than
$m_{\chi_1^0}$).  Using the DarkSUSY package \cite{Gondolo:2004sc}, we
calculated the thermal relic density of the LSP $\Omega_{\rm LSP}^{\rm
(thermal)}$.\footnote
{The Cases 1 and 2 are in the so-called ``co-annihilation region,''
while the Cases 3 and 4 are in the ``focus-point region'' and ``Higgs
funnel region,'' respectively.}
(We use $h=0.71$ \cite{WMAP}, where $h$ is the expansion rate of the
universe in units of $100\ {\rm km/sec/Mpc}$.)

\begin{table}[t]
  \begin{center}
    \begin{tabular}{lcccc}
      \hline\hline
      {} & {Case 1} & {Case 2} & {Case 3} & {Case 4}\\
      \hline
      $m_{1/2}$   
      & $300\ {\rm GeV}$  & $600\  {\rm GeV}$
      & $300\ {\rm GeV}$  & $1200\ {\rm GeV}$ \\
      $m_0$       
      & $141\  {\rm GeV}$ & $218\ {\rm GeV}$
      & $2397\ {\rm GeV}$ & $800\ {\rm GeV}$ \\
      $A_0$
      & $0$ & $0$ & $0$ & $0$ \\
      $\tan\beta$ & $30$ & $30$ & $30$ & $45$ \\
      $\mu_H$     
      & $389\ {\rm GeV}$ & $726\   {\rm GeV}$ 
      & $231\ {\rm GeV}$ & $-1315\ {\rm GeV}$ \\
      $m_{\chi_1^0}$
      & $117\ {\rm GeV}$ & $244\ {\rm GeV}$
      & $116\ {\rm GeV}$ & $509\ {\rm GeV}$ \\
      $\Omega_{\rm LSP}^{\rm (thermal)}h^2$ 
      & $0.111$ & $0.110$ & $0.106$ & $0.111$\\
      \hline\hline
    \end{tabular}
    \caption{mSUGRA parameters used in our analysis.}
    \label{table:mSUGRA}
  \end{center}
\end{table}

\section{Gravitino Decay}
\label{sec:decay}
\setcounter{equation}{0} 

In order to study the effects of the gravitino on the BBN, it is
important to understand the lifetime of the gravitino $\tau_{3/2}$ and
its decay modes.  Since the gravitino is the gauge field for the
supersymmetry, gravitino couples to the supercurrent and hence its
interaction is unambiguously determined.  Possible decay modes have,
however, model-dependence.  In the following, we discuss how the decay
rate and the branching ratios of the gravitino are calculated.

\subsection{Interaction and two-body processes}

We first briefly summarize the interactions of the gravitino.\footnote
{For details, see, for example, \cite{Moroi:1995fs}.}
Gravitino is the superpartner of the graviton, and it couples to the
supercurrent.  Thus, (relevant part of) the interaction of the
gravitino is given by
\begin{eqnarray}
  {\cal L}_{\rm int} &=& 
  - \frac{1}{8M_*} \sum_{G}
  \bar{\lambda}^{(G)} \gamma_5 \gamma^\mu 
  [ \gamma^\rho, \gamma^\sigma ] \psi_\mu F_{\rho\sigma}^{(G)}
  \nonumber \\ &&
  - \frac{1}{\sqrt{2}M_*} \sum_{C} 
  \left[ 
  \bar{\chi}^{(C)} \gamma^\mu \gamma^\nu P_L \psi_\mu 
  D_\nu \phi^{(C)} + {\rm h.c.} \right],
  \label{L_int}
\end{eqnarray}
where the sum over $G$ is for all the gauge multiplets (consisting of
the gauge field $A_\mu^{(G)}$ and the gaugino $\lambda^{(G)}$) while
$C$ for the chiral multiplets (consisting of the scalar boson
$\phi^{(C)}$ and the fermion $\chi^{(C)}$).  Here,
$F_{\rho\sigma}^{(G)}$ is the field strength for $A_\mu^{(G)}$, and
$D_\nu$ represents the covariant derivative.  In addition, $M_*\simeq
2.4\times 10^{18}\ {\rm GeV}$ is the reduced Planck scale.  As is
obvious from Eq.\ (\ref{L_int}), if we restrict ourselves to consider
the two-body final states, gravitino decays into some standard-model
particle and its superpartner.

From the Lagrangian given in Eq.\ (\ref{L_int}), we can read off the
vertex factors for the gravitino and calculate the (partial) decay
rates of the gravitino.  We first consider the decay processes with
two-body final states.  Then, the decay rate is expressed as
\begin{eqnarray}
    \Gamma_{\rm gauge} = 
    \frac{\beta_{\rm f} N_{\rm C}}{16\pi m_{3/2} M_*^2}
    \times
    \overline{\left| {\cal M} \right|^2},
    \label{Gamma_grav}
\end{eqnarray}
where ${\cal M}$ represents the Feynman amplitude for the decay
process (with $M_*=1$), and $N_{\rm C}$ is the color factor: $N_{\rm
C}=8$ for the process $\psi_\mu\rightarrow g\tilde{g}$, $N_{\rm C}=3$
for the processes with quark and squark final state, and $N_{\rm C}=1$
otherwise.  In addition, for the process $\psi_\mu\rightarrow AB$,
$\beta_{\rm f}$ is given by
\begin{eqnarray}
    \beta_{\rm f} = \frac{1}{m_{3/2}^2}
    \left[ m_{3/2}^4 - 2 m_{3/2}^2 (m_A^2 + m_B^2) 
        + (m_A^2 - m_B^2)^2 \right]^{1/2},
\end{eqnarray}
where $m_A$ and $m_B$ are masses of $A$ and $B$, respectively.

For the gravitino decay process into a gauge boson $V$ and a fermion
$\chi$, we define $p$, $q$, and $q'$ to be the momenta of $\psi_\mu$,
$V$, and $\chi$, respectively.  Then, we obtain
\begin{eqnarray}
    pq &=& \frac{1}{2} \left( m_{3/2}^2 + m_V^2 - m_\chi^2 \right),
    \label{pq}
    \\
    pq' &=& \frac{1}{2} \left( m_{3/2}^2 - m_V^2 + m_\chi^2 \right),
    \\
    qq' &=& \frac{1}{2} \left( m_{3/2}^2 - m_V^2 - m_\chi^2 \right),
    \label{qqdash}
\end{eqnarray}
where $m_V$ and $m_\chi$ are masses of $V$ and $\chi$, respectively.
With these quantities, for the process with a massless gauge field in
the final state (i.e., $\psi_\mu\rightarrow\gamma\chi_i^0$ and
$g\tilde{g}$), we obtain
\begin{eqnarray}
    \overline{\left| {\cal M} \right|^2}_{\gamma, g}
    = 
    \frac{2}{3} \left( C^{\rm (G)}_L C^{{\rm (G)}*}_L 
        + C^{\rm (G)}_R C^{{\rm (G)}*}_R \right)
    \left[ \frac{(pq)^2 (pq')}{m_{3/2}^2} + (pq) (qq') \right],
    \label{M2_gauge}
\end{eqnarray}
while with massive gauge field in the final state (i.e.,
$\psi_\mu\rightarrow Z\chi_i^0$ and $W^\pm\chi_i^\mp$) , we obtain
\begin{eqnarray}
    \overline{\left| {\cal M} \right|^2}_{W^\pm, Z} &=&
    \frac{2}{3} \left( C^{\rm (G)}_L C^{{\rm (G)}*}_L 
        + C^{\rm (G)}_R C^{{\rm (G)}*}_R \right)
    \left[ \frac{(pq)^2 (pq')}{m_{3/2}^2} + (pq) (qq') 
        - \frac{1}{2} m_V^2 (pq') 
    \right]
    \nonumber \\ &&
    - \left( C^{\rm (G)}_L C^{{\rm (G)}*}_R + C^{\rm (G)}_R C^{{\rm
    (G)}*}_L \right)
    m_{3/2} m_\chi m_V^2 
    \nonumber \\ &&
    + \frac{2}{3} \left( C^{\rm (G)}_L C^{{\rm (H)}*}_L
        - C^{\rm (G)}_R C^{{\rm (H)}*}_R + {\rm h.c.} \right)
    m_{3/2} 
    \left[ \frac{1}{2}(qq') + \frac{(pq) (pq')}{m_{3/2}^2}
    \right]
    \nonumber \\ &&
    + \left( C^{\rm (G)}_L C^{{\rm (H)}*}_R
        - C^{\rm (G)}_R C^{{\rm (H)}*}_L + {\rm h.c.} \right)
    m_\chi (pq)
    \nonumber \\ &&
    + \frac{2}{3} \left( C^{\rm (H)}_L C^{{\rm (H)}*}_L
        + C^{\rm (H)}_R C^{{\rm (H)}*}_R \right)
    \left[ 1 + \frac{(pq)^2}{2 m_{3/2}^2 m_V^2} \right]
    (pq')
    \nonumber \\ &&
    + \frac{2}{3} \left( C^{\rm (H)}_L C^{{\rm (H)}*}_R
        + C^{\rm (H)}_R C^{{\rm (H)}*}_L \right)
    \left[ 1 + \frac{(pq)^2}{2 m_{3/2}^2 m_V^2} \right]
    m_{3/2} m_\chi.
    \label{M2_wzgauge}
\end{eqnarray}
Here, $C_L^{\rm (G)}$, $C_R^{\rm (G)}$, $C_L^{\rm (H)}$, and $C_R^{\rm
(H)}$ are vertex factors.  These vertex factors for individual
processes are given in Appendix \ref{app:vertex}.  For the decay
processes with a scalar boson in the final state, we identify $p$,
$q$, and $q'$ to be the momenta of $\psi_\mu$, scalar boson $\phi$,
and fermion $\chi$, respectively.  Then, the products of the momenta
can be obtained from Eqs.\ (\ref{pq}) $-$ (\ref{qqdash}) by replacing
$m_V\rightarrow m_\phi$, with $m_\phi$ being the mass of the scalar
boson.  Then, we obtain
\begin{eqnarray}
    \overline{\left| {\cal M} \right|^2}_{\rm scalar} &=&
    \frac{1}{3} \left[ \frac{(pq)^2}{m_{3/2}^2} - m_\phi^2 \right]
    \nonumber \\ &&
    \left[ 
        \left( C^{\rm (C)}_L C^{{\rm (C)}*}_L 
            + C^{\rm (C)}_R C^{{\rm (C)}*}_R \right)
        (pq')
        + \left( C^{\rm (C)}_L C^{{\rm (C)}*}_R 
            + C^{\rm (C)}_R C^{{\rm (C)}*}_L \right)
        m_{3/2} m_\chi
    \right].
    \nonumber \\
    \label{M2_chiral}
\end{eqnarray}
The vertex factors $C_L^{\rm (C)}$ and $C_R^{\rm (C)}$ are also given
in Appendix \ref{app:vertex}.

\subsection{Three-body processes}

In this paper, we consider the case where the LSP is the lightest
neutralino $\chi_1^0$.  Then, the two-body process
$\psi_\mu\rightarrow\gamma\chi_1^0$ is always allowed, and the (total)
decay rate of the gravitino is determined by the two-body process(es).
In studying the effects of the gravitino on the BBN, however, it is
also important to understand the spectrum of the hadrons produced by
the decay of the gravitino.

In most of the cases, the number of the hadrons from the gravitino
decay is mostly determined by the two-body processes.  For example, if
the gravitino can decay into a superparticle other than the LSP, the
emitted superparticle decays into $\chi_1^0$.  In this secondary
decay process, sizable number of the hadrons may be produced.  In
addition, when the mass difference between the gravitino and the
lightest neutralino is larger than $m_Z$, the decay process
$\psi_\mu\rightarrow Z\chi_1^0$ becomes kinematically allowed.  In
this case, the decay of the $Z$-boson produces large amount of the
hadrons.  In most of the cases, the number of the hadrons produced
from those two-body processes is much larger than that from the three
body processes.  Then, the three-body processes are irrelevant for our
study.

However, in some case, precise determination of the hadron spectrum
requires the calculation of the three-body final state processes.  In
particular, the three-body processes become important if (i)
$m_{3/2}-m_{\chi_1^0}<m_Z$, and (ii) all the superparticles except
$\chi_1^0$ and sleptons are heavier than the gravitino.  Notice that,
when the condition (i) is satisfied, it may be the case that the only
possible two-body decay process is $\psi_\mu\rightarrow\gamma\chi_1^0$
and hence the hadrons are not produced by the two-body process.  In
some case, gravitino may also decay into lepton and slepton pair, but
the decays of the (s)leptons does not produce significant amount of
hadrons in our case.  Thus, we pay particular attention to the process
$\psi_\mu\rightarrow q\bar{q}\chi_1^0$ when
$m_{3/2}-m_{\chi_1^0}<m_Z$.

\begin{figure}[t]
    \begin{center}
        \centerline{{\vbox{\epsfxsize=0.6\textwidth\epsfbox{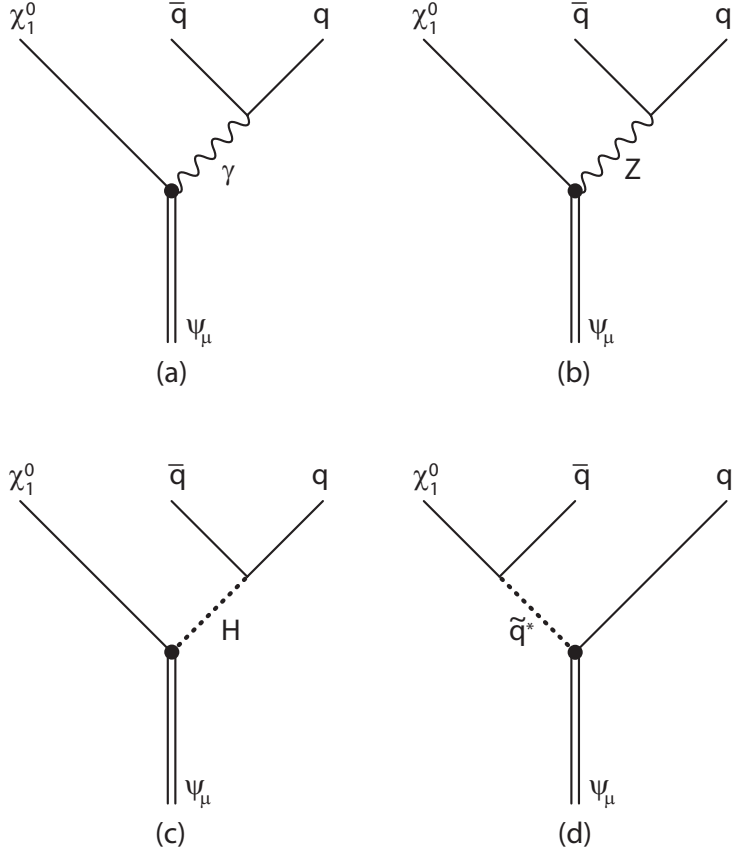}}}}
        \caption{Feynman diagrams for the process $\psi_\mu\rightarrow
        q\bar{q}\chi_1^0$.  The ``blobs'' are from the
        gravitino-supercurrent interaction.  For (d), there is also
        CP-conjugated diagram (with the replacements
        $q\leftrightarrow\bar{q}$ and
        $\tilde{q}^*\rightarrow\tilde{q}$).}
        \label{fig:Feyn3body}
    \end{center}
\end{figure}

\begin{figure}[t]
  \begin{center}
    \centerline{{\vbox{\epsfxsize=0.4\textwidth\epsfbox{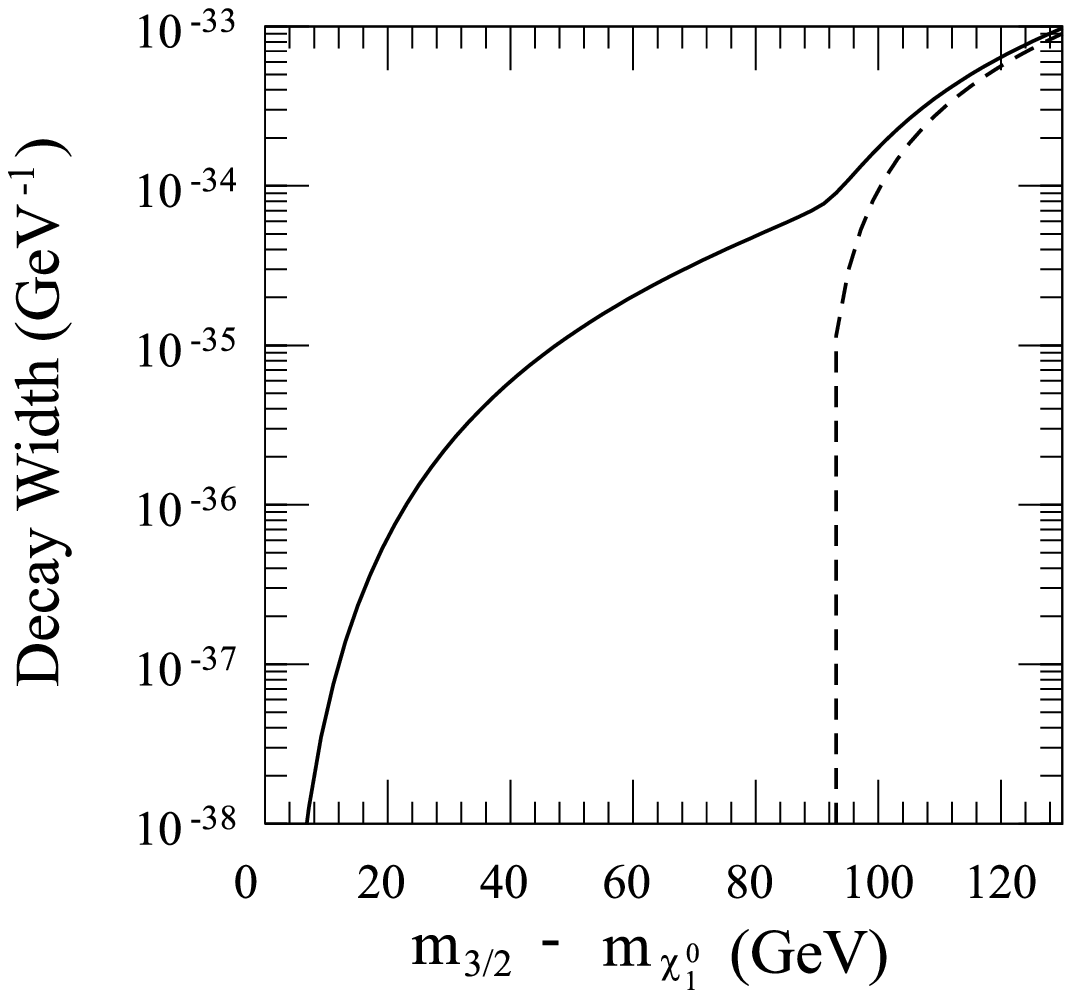}}}}
    \caption{Width for the process $\psi_\mu\rightarrow
    q\bar{q}\chi_1^0$ as a function of $m_{3/2}-m_{\chi^0_1}$ (solid
    line).  We adopt the mSUGRA parameters for the Case 1.  For
    comparison, the decay rate $\Gamma (\psi_\mu\rightarrow
    Z\chi_1^0)\times {\rm Br}(Z\rightarrow q\bar{q})$ is also shown in
    the dashed line.}
    \label{fig:width3body}
    \end{center}
\end{figure}

When the LSP is the neutralino, the relevant three-body processes are
induced by the diagrams shown in Fig.\ \ref{fig:Feyn3body}.  In our
study, we consider the effects of all the diagrams listed in Fig.\ 
\ref{fig:Feyn3body} and calculate the decay rate for the process
$\psi_\mu\rightarrow q\bar{q}\chi_1^0$.  To see the importance of the
3-body process, in Fig.\ \ref{fig:width3body}, we plot the ``3-body
hadronic decay width'' defined as\footnote
{As we discuss in the following, the three-body process becomes
important when the mass difference between the gravitino and the LSP
is smaller than $m_Z$.  Thus, the process $\psi_\mu\rightarrow
t\bar{t}\chi_1^0$ is not important for our analysis.}
\begin{eqnarray}
    \Gamma (\psi_\mu\rightarrow q\bar{q}\chi_1^0) &\equiv&
    \Gamma (\psi_\mu\rightarrow u\bar{u}\chi_1^0)
    + \Gamma (\psi_\mu\rightarrow d\bar{d}\chi_1^0)
    + \Gamma (\psi_\mu\rightarrow s\bar{s}\chi_1^0)
    \nonumber \\ &&
    + \Gamma (\psi_\mu\rightarrow c\bar{c}\chi_1^0)
    + \Gamma (\psi_\mu\rightarrow b\bar{b}\chi_1^0)
    + \Gamma (\psi_\mu\rightarrow t\bar{t}\chi_1^0).
\end{eqnarray}
In Fig.\ \ref{fig:width3body}, the MSSM parameters are taken to be
those for the Case 1.  In this case, the 3-body decay is induced
dominantly by the photon-mediated diagram.  When the mass difference
between the gravitino and the LSP becomes larger than $m_Z$, however,
$Z$-boson mediated contribution with $m_{q\bar{q}}\simeq m_Z$ becomes
the most important, where $m_{q\bar{q}}$ is the invariant mass of the
$q\bar{q}$ system.  In fact, such a process should rather be
classified into the 2-body process $\psi_\mu\rightarrow Z\chi_1^0$
followed by $Z\rightarrow q\bar{q}$.  To see this more explicitly, we
also plot the quantity $\Gamma(\psi_\mu\rightarrow Z\chi_1^0)\times
{\rm Br}(Z\rightarrow q\bar{q})$.  As one can see, $\Gamma
(\psi_\mu\rightarrow q\bar{q}\chi_1^0)$ is well approximated by
$\Gamma(\psi_\mu\rightarrow Z\chi_1^0)\times {\rm Br}(Z\rightarrow
q\bar{q})$ when the decay process $\psi_\mu\rightarrow Z\chi_1^0$
becomes kinematically allowed.

In our numerical study, we treat the process $\psi_\mu\rightarrow
q\bar{q}\chi_1^0$ in the following way:
\begin{itemize}
\item When $m_{3/2}-m_{\chi_1^0}<m_Z$, we calculate the Feynman
    diagrams shown in Fig.\ \ref{fig:Feyn3body} and calculate the
    decay rate $\Gamma(\psi_\mu\rightarrow q\bar{q}\chi_1^0)$.  (In
    this case, the decay process $\psi_\mu\rightarrow Z\chi_1^0$ is
    kinematically forbidden and hence is irrelevant.)
\item When $m_{3/2}-m_{\chi_1^0}>m_Z$, we approximate the hadronic
    decay rate induced by the diagrams in Fig.\ \ref{fig:Feyn3body} by
    $\Gamma(\psi_\mu\rightarrow Z\chi_1^0)\times {\rm Br}(Z\rightarrow
    q\bar{q})$.
\end{itemize}
With our treatment, the effect of the photon-mediated diagram (as well
as those from Figs.\ \ref{fig:Feyn3body}(c) and (d)) is neglected when
$m_{3/2}-m_{\chi_1^0}>m_Z$.  However, effect of such diagram is
subdominant since the process is mainly induced by the $Z$-boson
mediated diagram.

When $m_{3/2}-m_{\chi_1^0}>m_Z$, energy distribution of the quark and
anti-quark is easily calculated since the decay is dominated by the
process with $m_{q\bar{q}}\simeq m_Z$.  When
$m_{3/2}-m_{\chi_1^0}<m_Z$, on the contrary, $m_{q\bar{q}}$ has
broader distribution.  Thus, in this case, we numerically calculate
the differential decay rate
\begin{eqnarray*}
    \frac{d\Gamma(\psi_\mu\rightarrow q\bar{q}\chi_1^0)}
    {dm_{q\bar{q}}^2} 
\end{eqnarray*}
to obtain the energy distributions of the quarks and anti-quarks
emitted from the three-body processes.  (In the calculation of
$d\Gamma(\psi_\mu\rightarrow q\bar{q}\chi_1^0)/dm_{q\bar{q}}^2$, we
approximated that the final-state $q$ and $\bar{q}$ have isotropic
distribution in their center-of-mass frame.)  The hadron spectrum from
the three-body decay process is obtained by using this differential
decay rate.  When $m_{3/2}-m_{\chi_1^0}<m_Z$, the photon-mediated
diagram gives the dominant contribution while the effects of other
diagrams are almost irrelevant (unless $m_{3/2}-m_{\chi_1^0}$ is very
close to $m_Z$).  In Appendix \ref{app:3body}, we present the
approximated formula of the differential decay rate, only taking
account of the photon-mediated diagram.

\subsection{Lifetime and branching ratios of the gravitino}

\begin{figure}[t]
  \begin{center}
    \centerline{{\vbox{\epsfxsize=0.4\textwidth\epsfbox{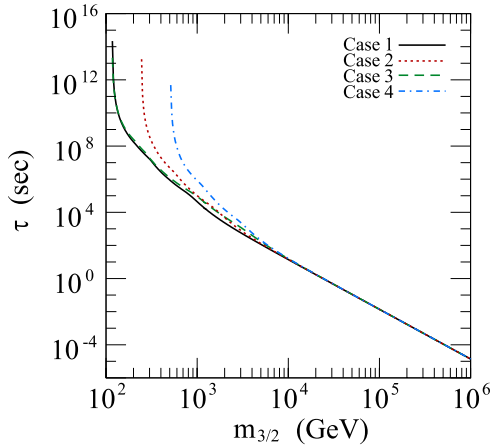}}}}
    \caption{Lifetime of the gravitino as a function of the gravitino
    mass.}
    \label{fig:lifetime}
    \end{center}
\end{figure}

Now we can quantitatively discuss the decay rates of the gravitino.
First, following the procedures discussed in the previous subsections,
we calculate the partial decay rates of the gravitino for all the
possible decay modes.  Adding all the contributions, we obtain the
lifetime of the gravitino:
\begin{eqnarray}
    \tau_{3/2} = \frac{1}{\Gamma (\psi_\mu\rightarrow {\rm all})}.
\end{eqnarray}
We calculate $\tau_{3/2}$ as a function of the gravitino mass for the
cases listed in Table \ref{table:mSUGRA}, and the results are shown in
Fig.\ \ref{fig:lifetime}.  Importantly, lifetime of the gravitino
becomes shorter as the gravitino becomes heavier.

As one can see, when the gravitino mass is smaller than a few TeV,
$\tau_{3/2}$ for the Case 4 is found to be longer than those for other
cases.  This is due to the fact that, for the Case 4, masses of the
MSSM particles are larger than other cases and hence the decay rates
of the gravitino in this case is suppressed.  When the gravitino is
much heavier than the MSSM particles, on the contrary, the lifetime of
the gravitino is insensitive to the mass spectrum of the MSSM
particles.

\begin{figure}[t]
  \begin{center}
    \centerline{{\vbox{\epsfxsize=0.7\textwidth\epsfbox{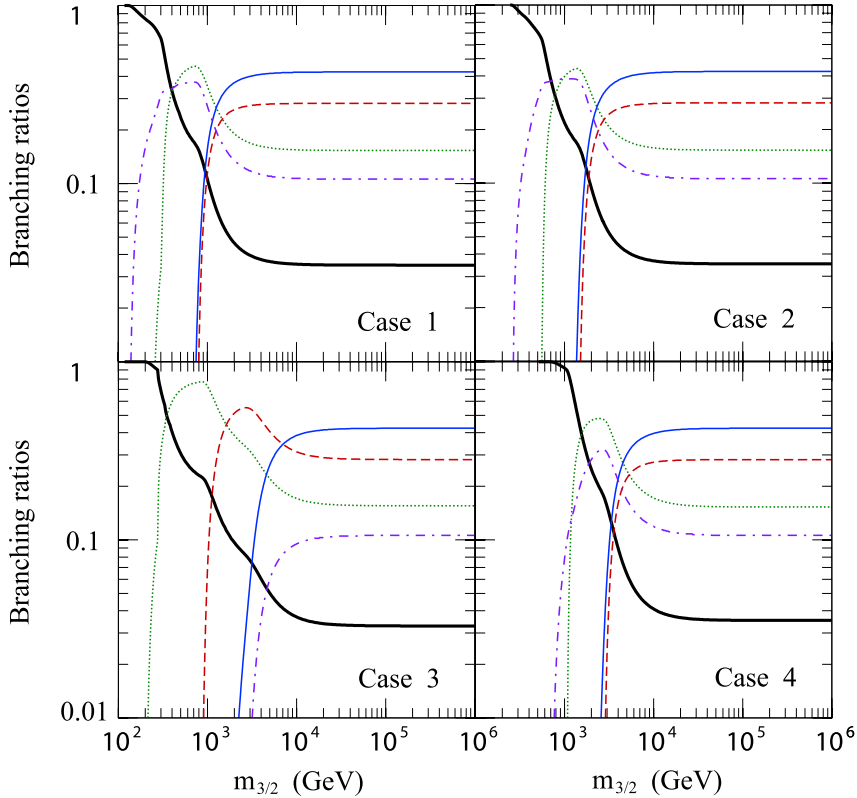}}}}
    \caption{Branching ratios of the decay of the gravitino as 
    functions of the gravitino mass.  The thick solid line is for the
    final states $\chi^0_1$ $+$ anything, dot-dashed line for
    lepton-slepton pairs, dotted line for $\chi^0_i$ ($i = 2-4$) or
    chargino $+$ anything, dashed line for gluon-gluino pair, and thin
    solid line for quark-squark pair final states.}
    \label{fig:br}
    \end{center}
\end{figure}

Importantly, when the gravitino is lighter than $\sim 20\ {\rm TeV}$,
$\tau_{3/2}$ becomes longer than $1\ {\rm sec}$ and hence the relic
gravitinos decay after the BBN starts.  Thus, in particular in this
case, significant constraints on the reheating temperature after the
inflation is expected.

Branching ratios of the decay process of the gravitino depends on the
model parameters.  To see this, for the Cases 1 $-$ 4, we plot the
branching ratios for various two-body final states in Fig.\ 
\ref{fig:br}.  As one can see, the branching ratios have sizable model
dependence when the gravitino mass is relatively small. This is
because, when the gravitino mass is small, decay rate of the gravitino
is sensitive to the mass spectrum of the superparticles.  When
$m_{3/2}$ becomes large enough, on the contrary, branching ratios
become insensitive to the model parameters.

\section{Secondary Decays and Hadronization}
\label{sec:hadronization}
\setcounter{equation}{0}

Although the gravitino primarily decays into a standard-model particle
and its superpartner (or into the $q\bar{q}\chi_1^0$ final state),
most of the daughter particles also decay with time scale much
shorter than the cosmological time scale.  In addition, all the
partons (i.e., quarks and gluon) are hadronized into mesons or
baryons.  These processes are important in the study of the BBN with
the primordial gravitinos and, in this section, we discuss these
issues.

The possible decay modes of the individual superparticles strongly
depend on the mass spectrum as well as on the mixing and coupling
parameters.  In order to systematically take account of all the
relevant decay processes, we use the ISAJET package \cite{ISAJET}
which automatically calculates the partial decay rates of all the
unstable particles.

In order to discuss the hadro-dissociations induced by the gravitino
decay, we should first calculate the spectra of the partons (i.e.,
$u$, $d$, $s$, $c$, $b$, $t$ and their anti-particles, and gluon).
There are two types of processes producing energetic partons: one is
the decay of the gravitino and the other is the subsequent decays of
the daughter particles.  Spectra of the partons of the first type are
directly calculated by using the partial decay rates of the gravitino
presented in the previous section.  (Here, the effects of the
``3-body'' processes are also included when
$m_{3/2}-m_{\chi_1^0}<m_Z$.)  In order to calculate the contribution
of the second type, we have to follow the decay chain of the unstable
particles.  In addition, the parton spectra should be calculated by
averaging over all the possible decay processes with the relevant
branching ratios of individual particles.  In our analysis, the decay
chain is followed by using the PYTHIA package \cite{PYTHIA} which
automatically take account of the decay processes of the unstable
particles (including the superparticles).

At the cosmic time relevant for the BBN, time scale for the
hadronization is much shorter than the time scale for the scattering
processes and hence all the partons are hadronized before scattering
off the background particles.  Thus, it is necessary to calculate the
spectrum of the mesons and baryons produced from the partons.  In
particular, for our analysis of the BBN-related processes, proton,
neutron, and charged pions play significant roles.  

\begin{figure}[t]
  \begin{center}
    \centerline{{\vbox{\epsfxsize=11.0cm\epsfbox{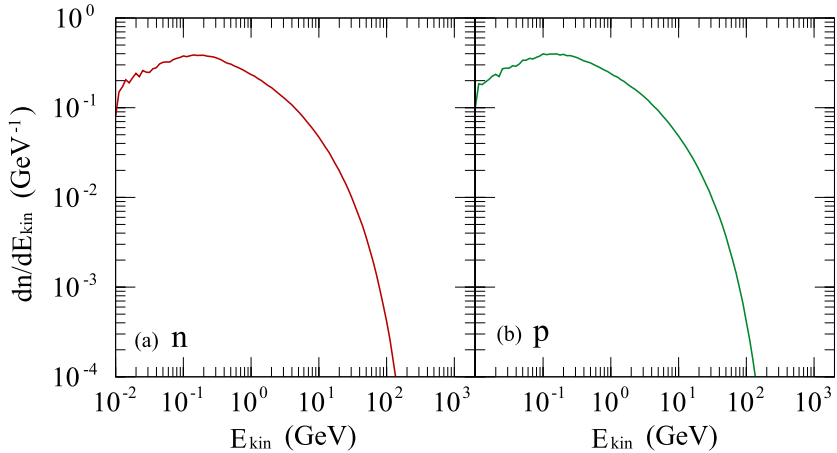}}}}
    \caption{Distributions of the nucleons 
    (i.e., (a) proton and (b) neutron) from the decay of a single
    gravitino as functions of the kinetic energy.  Here we take the
    mSUGRA parameters for the Case 1, and $m_{3/2}=1$TeV.}
    \label{fig:haddist}
    \end{center}
\end{figure}

In our analysis, the hadronization processes are dealt with the PYTHIA
package \cite{PYTHIA}; we have modified the PYTHIA package to include
the primary decay processes of the gravitino, then the subsequent
decay processes of the daughter particles (including the
superparticles) and the hadronization processes of the emitted partons
are automatically followed by the original PYTHIA algorithm.  In Fig.\ 
\ref{fig:haddist}, we plot the distribution functions of the proton
and neutron as functions of their kinetic energy (i.e., $E_{\rm
kin}=E-m_N$ with $m_N$ being the corresponding nucleon mass).  In
order to check the reliability of our estimate of the hadron spectrum,
we have performed an independent calculation using the ISAJET package
\cite{ISAJET}; we have also modified the ISAJET package to include the
decay processes of the gravitino and we calculated the hadron
spectrum.  We have checked that the difference between the two
calculations is within $10\ \%$ level.  In particular, for the region
with relatively large kinetic energy, which gives the most important
effects on the hadro-dissociation processes of the light elements,
difference is very small.

Before closing this section, we define one important parameter, which
is the (averaged) visible energy emitted from the gravitino decay.
Once a high energy particle with electro-magnetic interaction is
injected into the thermal bath, it induces the electro-magnetic shower
and, consequently, the photo-dissociation processes of the light
elements are induced by the energetic photons in the shower.  The
event rates of the photo-dissociation processes (for a fixed
background temperature) are mostly determined by the total amount of
the ``visible'' energy injected into the thermal bath \cite{KawMor}.
Since we consider the case where the $R$-parity is conserved, some
fraction of the energy is always carried away by the LSP for the decay
process of the gravitino.  Taking account of such effect, we calculate
the averaged visible energy emitted by a single decay of the
gravitino:\footnote
{For unstable leptons and mesons (in particular, pions), we checked
that their lifetimes are shorter than their mean free time.
Thus, in the calculation of the visible energy, we treated them as
unstable particles and hence the energy carries away by the neutrinos
are not included in $E_{\rm vis}$.}
\begin{eqnarray}
  E_{\rm vis} = m_{3/2} - \langle E_{\chi_1^0} \rangle 
  - \langle E_\nu \rangle,
\end{eqnarray}
where the second and third terms of the right-hand side are the
(averaged) energy carried away by the LSP and the neutrinos,
respectively.  $E_{\rm vis}$ is used for the calculations of the
photo-dissociation rates.

\section{Light-Element Abundances}
\label{sec:abundance}
\setcounter{equation}{0}

\subsection{Theoretical calculation}

Now, we explain how we calculate the light-element abundances.  In
order to set a bound on the reheating temperature after the inflation,
we assume that the gravitinos are produced by the scattering processes
of the thermal particles.  Using the thermally averaged gravitino
production cross section given in \cite{Bolz:2000fu}, the ``yield
variable'' of the gravitino, which is defined as
$Y_{3/2}\equiv\frac{n_{3/2}}{s}$, is given by \cite{Kawasaki:2004qu}
\begin{eqnarray}
    \label{eq:Yx-new}
    Y_{3/2} &\simeq& 
    1.9 \times 10^{-12}
    \nonumber \\ &&
    \times \left( \frac{T_{\rm R}}{10^{10}\ {\rm GeV}} \right)
    \left[ 1 
        + 0.045 \ln \left( \frac{T_{\rm R}}{10^{10}\ {\rm GeV}} 
        \right) \right]
    \left[ 1 
        - 0.028 \ln \left( \frac{T_{\rm R}}{10^{10}\ {\rm GeV}} 
        \right) \right],
    \nonumber \\
\label{Ygrav}
\end{eqnarray}
where $n_{3/2}$ is the number density of the gravitino while
$s=\frac{2\pi^2}{45}g_{*S}(T)T^3$ is the entropy density with
$g_{*S}(T)$ being the effective number of the massless degrees of
freedom at the temperature $T$, and the reheating temperature is
defined as\footnote
{Strictly speaking, the ``reheating temperature'' corresponds to the
maximal temperature of the last radiation dominated era.  Thus, if
some scalar field $\phi$ other than the inflaton once dominates the
universe after the inflation, the reheating temperature here is given
by the same expression as Eq.\ (\ref{T_R}) with $\Gamma_{\rm inf}$
being replaced by the decay rate of $\phi$.  One of the examples of
such scalar fields is the curvaton \cite{curvaton} which provides a
new origin of the cosmic density perturbations.}
\begin{eqnarray}
  T_{\rm R} \equiv 
  \left( 
  \frac{10}{g_* \pi^2} M_*^2 \Gamma_{\rm inf}^2 
  \right)^{1/4},
  \label{T_R}
\end{eqnarray}
with $\Gamma_{\rm inf}$ being the decay rate of the inflaton.

Once produced, the gravitinos decay with very long lifetime.  In
particular, if the gravitino mass is smaller than $\sim 20\ {\rm
TeV}$, gravitinos decay after the BBN starts and hence the
light-element abundances may be significantly affected.  In order to
study the BBN processes with unstable gravitino, our study proceeds as
follows:
\begin{enumerate}
\item MSSM parameters are determined for one of the mSUGRA points
  given in Table \ref{table:mSUGRA}.  Then, all the mass eigenvalues
  and mixing parameters are calculated.
\item Using the parameters given above, we calculate partial decay
    rates of the gravitino for all the kinematically allowed 2-body
    processes.  When $m_{3/2}-m_{\chi_1^0}<m_Z$, we also calculate
    $\Gamma(\psi_\mu\rightarrow q\bar{q}\chi_1^0)$.
\item We perform a Monte-Carlo analysis using the branching ratios of
    the gravitino to calculate the energy distribution of the hadrons
    produced by the decay of the gravitino.  As explained in the
    previous sections, the decay chain of the decay products
    (including the superparticles) and the hadronizations of the
    emitted partons are followed by the modified PYTHIA code.
    Simultaneously, we calculate the (averaged) emitted visible energy
    from the decay of the gravitino.
\item For a given reheating temperature $T_{\rm R}$, we calculate the
    abundance of the thermally produced gravitino using Eq.\
    (\ref{Ygrav}).
\item We calculate the light-element abundances, taking account of the
    decay of the thermally produced gravitinos.  (Standard reactions
    of the light elements are also included.)  We use the
    baryon-to-photon ratio suggested by the WMAP~\cite{WMAP}:
    \begin{eqnarray}
        \eta = (6.1 \pm 0.3) \times 10^{-10},
    \end{eqnarray}
    at the 1$\sigma$ level.  (Here we enlarged the lower error bar
    from 0.2 to 0.3 since the Monte-Carlo simulation is easier if the
    error bar is symmetric.  This does not significantly change the
    resultant constraints.)  The baryon-to-photon ratio is related to
    the density parameter of the baryon as $\Omega_Bh^2=3.67\times
    10^7\eta$.
\item Since the event rates of the non-standard processes induced by
    the gravitino decay are proportional to the abundance of the
    primordial gravitinos, deviations of the light element abundances
    from the standard BBN results become larger as the reheating
    temperature becomes higher.  The resultant light element
    abundances are compared with the observational constraints and
    upper bound on the reheating temperature is obtained.
\end{enumerate}

Although the details of the analysis of the photo- and
hadro-dissociation processes and $p\leftrightarrow n$ interchange are
explained in
\cite{KawMor,Holtmann:1998gd,Kawasaki:2000qr,Kawasaki:2004qu}, we
briefly summarize several important points.  Once energetic hadrons
are emitted into the thermal bath in the early universe, they induce
the hadronic shower and energetic particles in the shower cause
hadro-dissociation processes.  In addition, released visible energy
from the gravitino decay eventually goes into the form of radiation
which cause electro-magnetic shower.  Energetic photons in the shower
cause photo-dissociation processes.  Furthermore, when the cosmic
temperature is relatively high ($T \gtrsim$ 0.1 MeV),
$p\leftrightarrow n$ inter-converting processes by the nucleons and
the charged pions become significant.

When the cosmic temperature is higher than 0.3 MeV, the
$p\leftrightarrow n$ inter-converting processes induced by the charged
pions (i.e., $p+\pi^-\rightarrow n+\pi^0$ and $n+\pi^+\rightarrow
p+\pi^0$) are the most important.  Since the resultant $^4$He
abundance is sensitive to the $n/p$ ratio, such inter-converting
processes affects the $^4$He abundance.

\begin{figure}[t]
    \begin{center}
    \centerline{{\vbox{\epsfxsize=8.0cm\epsfbox{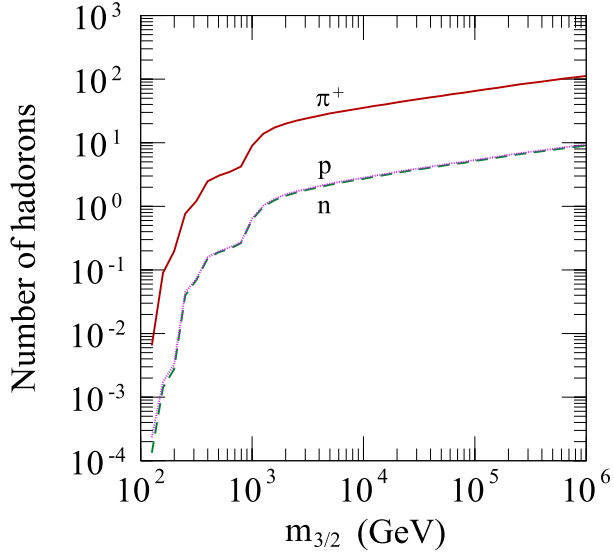}}}}
        \caption{Number of hadrons($\pi$, $p$, $n$) produced by 
        the decay of single gravitino for the Case 1.  (The 
        number of $\pi^-$ is the same as that of $\pi^+$ within the
        expected error of the Monte-Carlo analysis.)}
        \label{fig:n_pi}
    \end{center}
\end{figure}

Since the charged pions are expected to be stopped in the thermal bath
before inter-converting the background nucleons, we need to know only
the total numbers of $\pi^+$ and $\pi^-$ produced by the decay of the
gravitino.  The number of pions produced by the decay of the single
gravitino is plotted in Fig.\ \ref{fig:n_pi} as a function of the
gravitino mass for the Case 1.  As one can see, the number of the
pions increases as the gravitino mass becomes larger. In addition,
when the gravitino mass is smaller than $\sim 1\ {\rm TeV}$, partial
decay rates of the gravitino have significant dependence on $m_{3/2}$
because the gravitino mass becomes close to the masses of the MSSM
superparticles in this region; consequently, number of pion has strong
dependence on $m_{3/2}$.  In the figure, we also plotted the number of
the proton and the neutron produced by the decay of the gravitino.

As the lifetime of the gravitino becomes longer, it is likely that
most of the thermally produced gravitinos decay after $^4$He and other
light elements (like D, T, $^3$He, and so on) are synthesized.  Then,
the hadro- and photo-dissociation processes may significantly change
the light-element abundances.

When $10^2\ {\rm sec}\lesssim\tau_{3/2}\lesssim 10^7\ {\rm sec}$,
hadro-dissociation processes of the light elements are important
while, for longer lifetime, photo-dissociation processes give more
significant constraints.  In particular, since the number density
$^4$He is much larger than those of other light elements, dissociation
of $^4$He may significantly change the abundances of D and $^3$He.  In
addition, non-thermally produced T and $^3$He may scatter off the
background $^4$He to produce $^6$Li via the processes ${\rm T}+{\rm
^4He}\rightarrow {\rm ^6Li}+p$ and ${\rm ^3He}+{\rm ^4He}\rightarrow
{\rm ^6Li}+n$
\cite{Jedamzik:1999di,Kawasaki:2000qr,Kawasaki:2004yh,Kawasaki:2004qu}.
Since there is very stringent upper bound on the primordial abundance
of $^6$Li, such non-thermal production process of $^6$Li gives
significant constraint on the reheating temperature.

\subsection{Observational constraints}

In order to derive a bound on the reheating temperature, we compare
the theoretical results of the light-element abundances with the
observational constraints.  The observational constraints we use are
summarized below.  Since there are uncertainties in the constraints,
we adopt several different bounds on the primordial abundances of the
light elements in some case. The errors of the following observational
values are at $1\sigma$ level unless otherwise stated. When we adopt
both the statistical and systematic errors, we add them in quadrature.

For the ${\rm D}$ abundance, we use constraints obtained from the
measurements in the high redshift QSO absorption systems \cite{ObsD}.
Here, we consider two constraints; one is the ``averaged'' constraint
\begin{equation}
  (n_{\rm D}/n_{\rm H})^{\rm obs} 
  = (2.78^{+0.44}_{-0.38}) \times 10^{-5},
  \label{lowd}
\end{equation}
while the other is the ``conservative'' one, which is the highest
value of $n_{\rm D}/n_{\rm H}$ among the results listed in
\cite{ObsD}:
\begin{equation}
    \label{highd}
    (n_{\rm D}/n_{\rm H})^{\rm obs} 
    = (3.98^{+0.59}_{-0.67}) \times 10^{-5}.
\end{equation}
(Here and hereafter, the superscript ``obs'' is used for the
primordial values inferred by the observations.)

Abundance of ${\rm ^3He}$ may significantly change during the
evolution of the universe from the BBN epoch to the present epoch.
Thus, for ${\rm ^3He}$, it is not easy to observationally determine
its primordial value.  In our analysis, we do not rely on any detailed
model of chemical evolution to derive bound on the primordial
abundance of ${\rm ^3He}$.  Instead, we only use the fact that ${\rm
D}$ is more fragile than ${\rm ^3He}$.  Then, we expect that the ratio
$r_{3,2}\equiv n_{\rm ^3He}/n_{\rm D}$ does not decrease with
time~\cite{Sigl:1995kk,Kawasaki:2004yh,Kawasaki:2004qu,Ellis:2005ii}.
The solar-system value of ${\rm ^3He}$-to-${\rm D}$ ratio is measured
as \cite{Geiss93}
\begin{equation}
    r_{3,2}^{\odot} = 0.59 \pm 0.54 ~~ (2\sigma).
    \label{He3/D}
\end{equation}
Thus, we obtain the upper bound on 
the primordial $^3$He to D ratio
\begin{equation}
    \label{upper_bound32}
    r_{3,2}^{\rm obs} \le r_{3,2}^{\odot}.
\end{equation}

For the primordial abundance of $^4$He, we use the constraints from
the recombination lines from the low metallicity extragalactic HII
regions.  Taking into account the fact that several groups
independently derived bounds on the mass fraction of $^4$He, we derive
upper bound on $T_{\rm R}$ with the following three different bounds:
the first one is based on the analysis by Fields and Olive
\cite{Fields:1998gv}
\begin{equation}
    Y^{\rm obs}({\rm FO}) = 0.238 \pm (0.002)_{\rm stat} \pm
    (0.005)_{\rm syst}, 
    \label{Y_FO}
\end{equation}
where the first and second errors are the statistical and systematic
ones, respectively, the second is obtained by Izotov and Thuan
\cite{Izotov:2003xn}:
\begin{equation}
    Y^{\rm obs}({\rm IT}) = 0.242 \pm (0.002)_{\rm stat} (\pm
    (0.005)_{\rm syst}),
    \label{Y_IT}
\end{equation}
where we have added the systematic errors following Refs.\ 
\cite{Olive:1994fe,Olive:1996zu,Izotov:mi}, and the last one is by
Olive and Skillman~\cite{Olive:2004kq}:
\begin{equation}
    Y^{\rm obs}({\rm OS}) = 0.249 \pm 0.009,
    \label{Y_OS}
\end{equation}
where the error includes both the statistical and systematic ones.
With these three constraints, we will discuss how the upper bound on
$T_{\rm R}$ changes as we adopt different value of $Y^{\rm obs}$.

The primordial value of the $^{7}{\rm Li}$ abundance is observed in
old Pop II halo stars. Typically $n_{^7{\rm Li}}/n_{\rm H}$ is
$O(10^{-10})$.  In \cite{RyanLi7}, it was reported that
\begin{eqnarray}
  \left( n_{^7{\rm Li}}/n_{\rm H} \right)^{\rm obs} = 
  1.23^{+0.68}_{-0.32} \times 10^{-10},
  \label{Li7_Ryan}
\end{eqnarray}
while, recently, relatively higher value of the $^{7}{\rm Li}$
abundance was also reported \cite{Bonifacio:2002yx}:
\begin{eqnarray}
    \log_{10} [ \left(n_{^7{\rm Li}}/n_{\rm H}\right)^{\rm obs}] =
    -9.66\pm (0.056)_{\rm stat} \pm (0.06)_{\rm syst}.
    \label{Li7_Bonifacio}
\end{eqnarray}
Here, ${\rm ^7Li}$ abundances given in (\ref{Li7_Ryan}) and
(\ref{Li7_Bonifacio}) differ by the factor $\sim 2$, and we expect
that there is still some large uncertainty for the observational
values of $n_{^7{\rm Li}}/n_{\rm H}$.  Here, we adopt the constraint
given in \cite{Bonifacio:2002yx} with an additional large systematic
error, considering the possibilities of the increase by the cosmic-ray
spallation of the C, N, O and so on, and the decrease by the depletion
by the convection in the stars \cite{factor-of-two}
\begin{eqnarray}
  \label{eq:li7}
  \log_{10}
  \left[ \left(n_{^7{\rm Li}}/n_{\rm H}\right)^{\rm obs} \right]
  =-9.66 \pm (0.056)_{\rm stat} \pm (0.300)_{\rm add}.
\end{eqnarray}
The linear-scale value is given by $(n_{^7{\rm Li}}/n_{\rm H})^{\rm
obs} = (0.54 - 8.92) \times 10^{-10}$ at the $2\sigma$ level.  In
deriving the upper bound on $T_{\rm R}$, we use the constraint
(\ref{eq:li7}).

Usually the abundance of $^6{\rm Li}$ is measured as a ratio of
$^6{\rm Li}$ and $^7{\rm Li}$ in the old Pop II halo stars; at the
2$\sigma$ level, $(n_{\rm ^6Li} / n_{^7{\rm Li}})^{\rm halo} = 0.05
\pm 0.02$ \cite{li6_obs}.  The primordial value is expected to be
smaller than this value because it is likely that the the cosmic-ray
spallation has produced additional $^6{\rm Li}$ after the BBN
\cite{li6metal,li6LSTC,li6FO}. By adopting the milder value of the
constraint on $n_{^7{\rm Li}}/n_{\rm H} $ given in Eq.~(\ref{eq:li7}),
we get the upper bound on the primordial value of $n_{\rm ^6Li} /
n_{\rm H}$ at the 2$\sigma$ level,
\begin{eqnarray}
    (n_{\rm ^6Li} / n_{\rm H})^{\rm obs} <
    ( 1.10^{+ 5.14}_{-0.94} ) \times 10^{-11} ~~ (2 \sigma).
    \label{eq:obs6_upp}
\end{eqnarray}
We use this value as the upper bound on $n_{\rm ^6Li} / n_{\rm H}$
except in Section \ref{sec:li7crisis}.

\section{Numerical Results}
\setcounter{equation}{0}
\label{sec:results}

\subsection{Upper bound on $T_{\rm R}$}

Now, we are at the position to show our numerical results.  As
discussed in the previous sections, we calculate the light element
abundances as functions of the gravitino mass, other MSSM parameters,
and the reheating temperature $T_{\rm R}$.  Then, we compare the
theoretical prediction with the observations.  In order to
systematically derive the upper bound, we calculate the $\chi^2$
variable defined as
\begin{eqnarray}
\label{eq:chi2_1}
    \chi^2_i = 
    \frac{ ( \bar{x}_i^{\rm th} - \bar{x}_i^{\rm obs} )^2}
    { (\sigma_i^{\rm th})^2 + (\sigma_i^{\rm obs})^2 }
    ~~ {\rm for}\ x_i = (n_{\rm D}/n_{\rm H}), Y, 
\end{eqnarray}
where $\bar{x}_i^{\rm th}$ and $\bar{x}_i^{\rm obs}$ are the center
values of $x_i$ determined from the theoretical calculation and
observations, while $\sigma_i^{\rm th}$ and $\sigma_i^{\rm obs}$ are
their errors, respectively.  In our analysis, $(\sigma_i^{\rm th})^2$
is calculated by the Monte-Carlo analysis.  For $x_i=r_{3,2}$ $(n_{\rm
^6Li}/n_{\rm H})$ and $\log_{10}[(n_{\rm ^7Li}/n_{\rm H})]$ we only
use the upper bound, and we define $\chi^2_i$ as\footnote
{For $n_{\rm ^7Li}$, we only use the upper bound  since we could not
include one of the non-thermal production process of $n_{\rm ^7Li}$,
because of the lack of the experimental data.  For details, see
\cite{Kawasaki:2004qu}, and also the next subsection.}
\begin{eqnarray}
\label{eq:chi2_2}
    \chi^2_i =
    \left\{ 
        \begin{array}{ll}
            {\displaystyle{
            \frac{ ( \bar{x}_i^{\rm th} - \bar{x}_i^{\rm obs} )^2}
            { (\sigma_i^{\rm th})^2 + (\sigma_i^{\rm obs})^2 } } }
            &:\bar{x}_i^{\rm th} < \bar{x}_i^{\rm obs} 
            \\ \\
            0 &:{\rm otherwise}
        \end{array} 
    \right.
    {\rm for}\ x_i = r_{3,2},~(n_{\rm ^6Li}/n_{\rm H}),
    \log_{10}[(n_{\rm ^7Li}/n_{\rm H})]. 
\end{eqnarray}
With these quantities, we derive 95\ \% level constraints which
correspond to $\chi^2_i=3.84$ for $x_i= (n_{\rm D}/n_{\rm H})$ and
$Y$, and $\chi^2_i=2.71$ for $x_i=r_{3,2}$, ($n_{\rm ^6Li}/n_{\rm H}$)
and $\log_{10}[(n_{\rm ^7Li}/n_{\rm H})]$.  (For details, see
\cite{Kawasaki:2004qu}.)

\begin{figure}[t]
    \begin{center}
        \centerline{{\vbox{\epsfxsize=0.6\textwidth\epsfbox{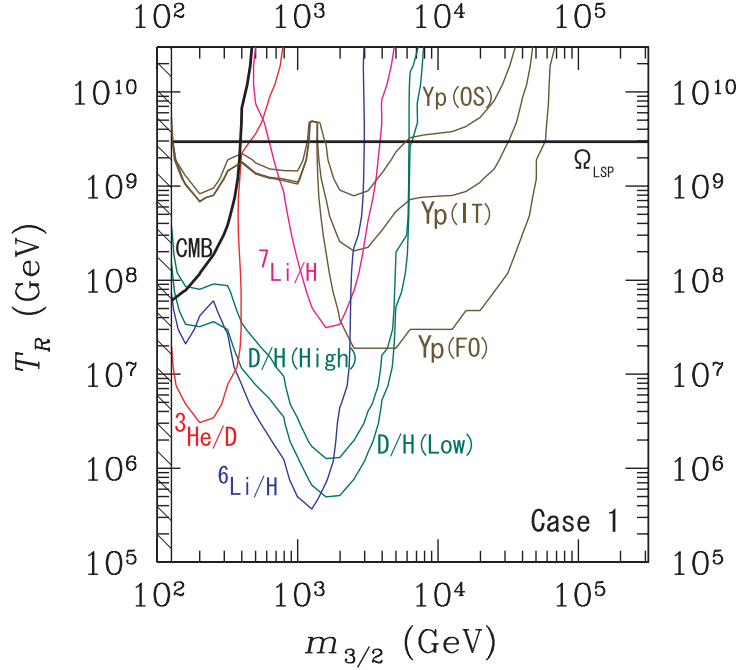}}}}
        \caption{Upper bound on the reheating temperature for
        the Case 1 as a function of the gravitino mass.}
        \label{fig:mtr1}
    \end{center}
\end{figure}

\begin{figure}[t]
    \begin{center}
        \centerline{{\vbox{\epsfxsize=0.6\textwidth\epsfbox{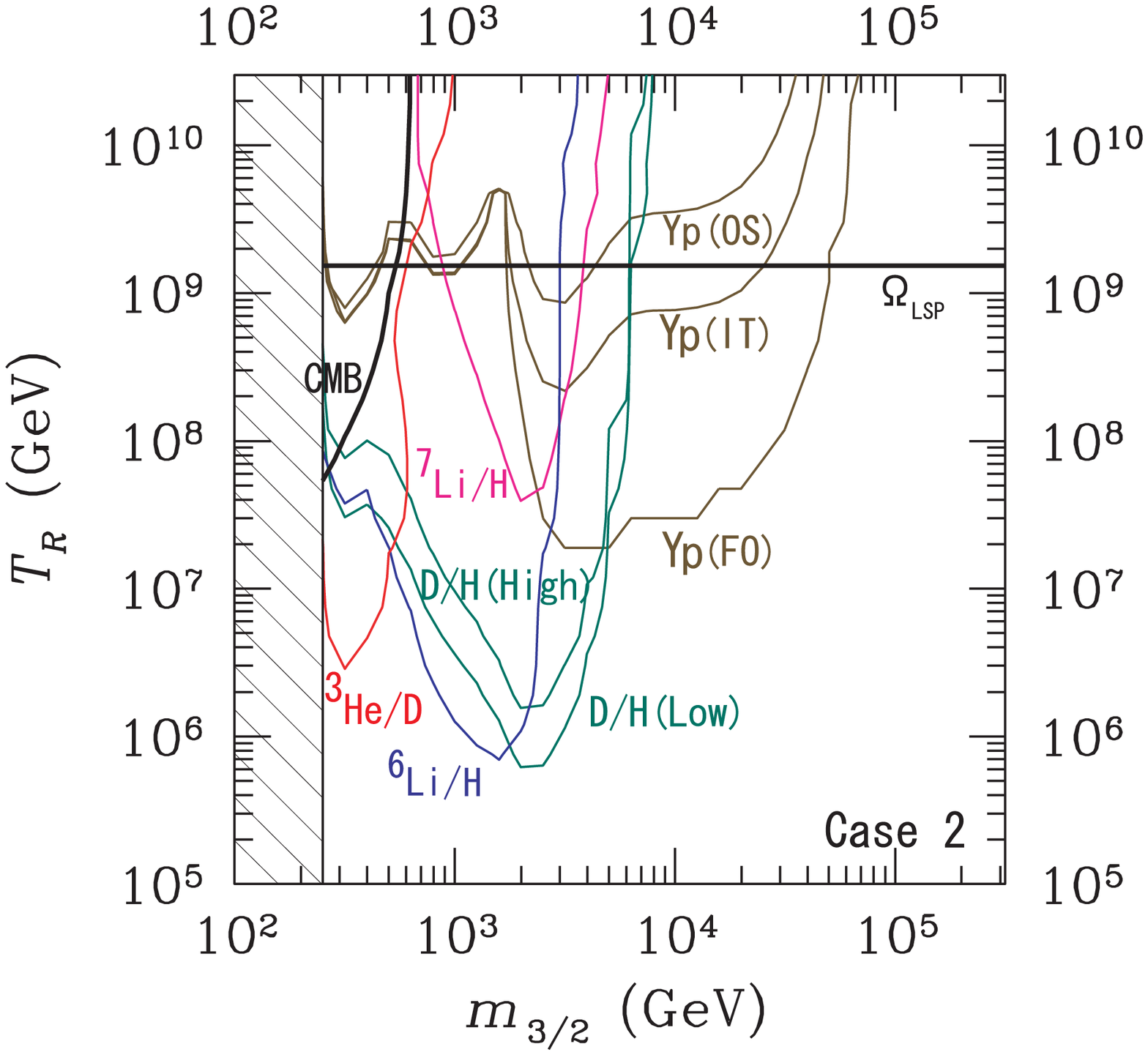}}}}
        \caption{Same as Fig.\ \ref{fig:mtr1} except for the MSSM
        parameters are evaluated for the Case 2.}
        \label{fig:mtr2}
    \end{center}
\end{figure}

\begin{figure}[t]
    \begin{center}
        \centerline{{\vbox{\epsfxsize=0.6\textwidth\epsfbox{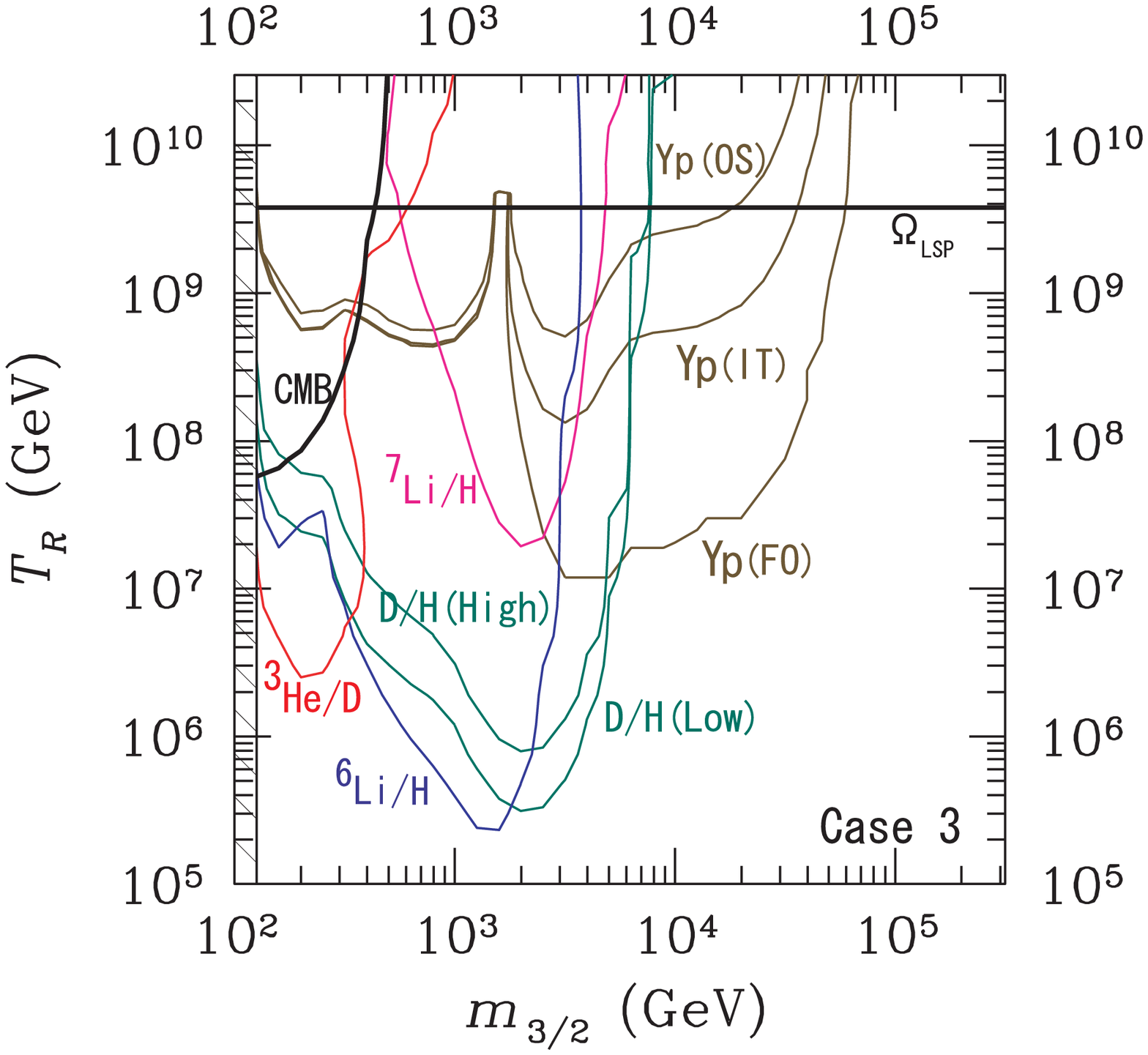}}}}
        \caption{Same as Fig.\ \ref{fig:mtr1} except for the MSSM
        parameters are evaluated for the Case 3.}
        \label{fig:mtr3}
    \end{center}
\end{figure}

\begin{figure}[t]
    \begin{center}
        \centerline{{\vbox{\epsfxsize=0.6\textwidth\epsfbox{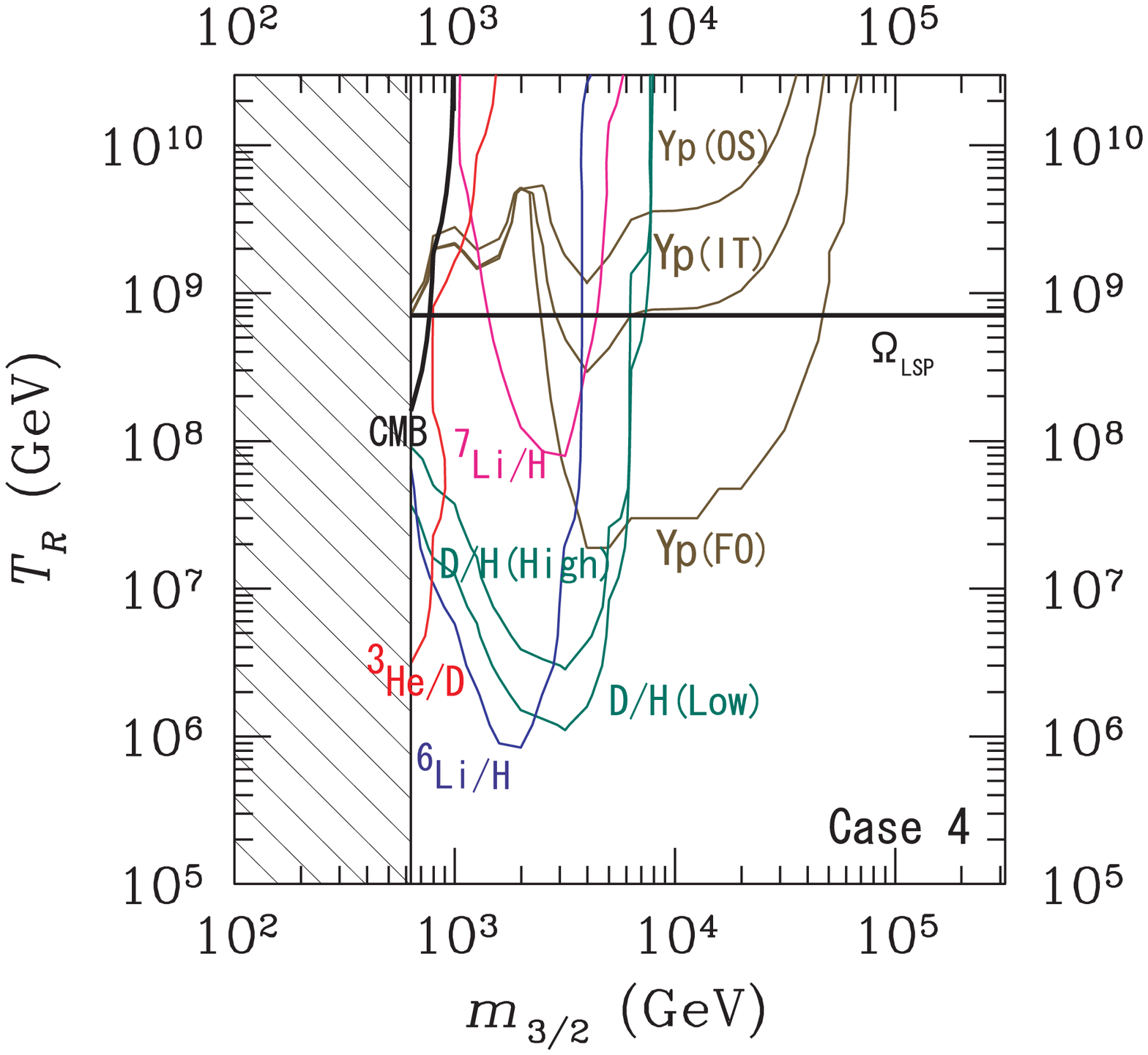}}}}
        \caption{Same as Fig.\ \ref{fig:mtr1} except for the MSSM
        parameters are evaluated for the Case 4.}
        \label{fig:mtr4}
    \end{center}
\end{figure}

In Figs.\ \ref{fig:mtr1} $-$ \ref{fig:mtr4}, we show the upper bounds
on the reheating temperature.  For ${\rm D}$, we considered the
observational constraints (\ref{lowd}) and (\ref{highd}) to see how
the upper bound depends on the bound on ${\rm D}$.  For ${\rm ^3He}$,
${\rm ^7Li}$, and ${\rm ^6Li}$, we use (\ref{He3/D}), (\ref{eq:li7}),
and (\ref{eq:obs6_upp}), respectively.  For ${\rm ^4He}$, we consider
three cases (\ref{Y_FO}) (FO), (\ref{Y_IT}) (IT), (\ref{Y_OS}) (OS),
since the upper bound on $T_{\rm R}$ for the case of relatively heavy
gravitino is sensitive to the observational constraint on the
abundance of ${\rm ^4He}$.  In deriving Figs.\ \ref{fig:mtr1} $-$
\ref{fig:mtr4}, the MSSM parameters are determined by using the mSUGRA
parameters given in Table \ref{table:mSUGRA}.  In addition, the
lifetime of the gravitino as well as its branching ratios are
calculated using the MSSM mass spectrum obtained from these
parameters.  In this analysis, we concentrated on the case where the
gravitino is unstable.  In the figures, we shaded the region where
$m_{3/2}<m_{\chi^0_1}$.

Although we have considered four different cases with different mass
spectrum of the MSSM particles, the qualitative behavior of the
constraints are quite insensitive to the choice of underlying
parameters.  When the gravitino mass is larger than a few TeV, most of
the primordial gravitinos decay at very early stage of the BBN.  In
this case, in addition, photo- and hadro-dissociations are
ineffective.  Then, overproduction of ${\rm ^4He}$ due to the
$p\leftrightarrow n$ conversion becomes the most important.  For the
observational constraints on the mass fraction of ${\rm ^4He}$, we
consider three different observational results given in Eqs.\ 
(\ref{Y_FO}) $-$ (\ref{Y_OS}).  As one can see, the upper bound on
$T_{\rm R}$ in this case is sensitive to the observational constraint
on the primordial abundance of ${\rm ^4He}$; for the case of
$m_{3/2}=10\ {\rm TeV}$, for example, $T_{\rm R}$ is required to be
lower than $3\times 10^7\ {\rm GeV}$ if we use the lowest value of $Y$
given in Eq.\ (\ref{Y_FO}) while, with the highest value given in
Eq.\ (\ref{Y_OS}), the upper bound on the reheating temperature
becomes as large as $4\times 10^9\ {\rm GeV}$.

When $400\ {\rm GeV}\lesssim m_{3/2}\lesssim 5\ {\rm TeV}$, gravitinos
decay when the cosmic temperature is $1\ {\rm keV}$ $-$ $100\ {\rm
keV}$.  In this case, hadro-dissociation gives the most stringent
constraints; in particular, the overproductions of ${\rm D}$ and ${\rm
^6Li}$ become important.  Furthermore, when the gravitino mass is
relatively light ($m_{3/2}\lesssim 400\ {\rm GeV}$), the most
stringent constraint is from the ratio ${\rm ^3He}/{\rm D}$ which may
be significantly changed by the photo-dissociation of ${\rm ^4He}$.

It should be noted that, even when the gravitino cannot directly decay
into colored particles (i.e., the squarks, gluino, and their
superpartners) due to the kinematical reason, the reheating
temperature may still be stringently constrained from the
hadro-dissociation processes.  This is due to the fact that some of
the non-colored decay products (in particular, the weak bosons $W^\pm$
and $Z$ as well as some of the superparticles) produce hadrons when
they decay.  In particular, when the mass difference
$m_{3/2}-m_{\chi^0_1}$ is larger than $m_Z$, hadro-dissociations
become important since sizable amount of hadrons are produced by the
process $\psi_\mu\rightarrow Z\chi^0_1$.

\begin{table}[t]
    \begin{center}
        \begin{tabular}{ccccc}
            \hline\hline
            $m_{3/2}$ & {Case 1} & {Case 2} & {Case 3} & {Case 4}\\
            \hline
            300\ GeV 
            & $6 \times 10^{6}\ {\rm GeV}$
            & $3 \times 10^{6}\ {\rm GeV}$
            & $4 \times 10^{6}\ {\rm GeV}$
            & $-$ \\
            1\ TeV 
            & $5 \times 10^{5}\ {\rm GeV}$
            & $1 \times 10^{6}\ {\rm GeV}$
            & $4 \times 10^{5}\ {\rm GeV}$
            & $6 \times 10^{6}\ {\rm GeV}$\\
            3\ TeV 
            & $1 \times 10^{6}\ {\rm GeV}$
            & $1 \times 10^{6}\ {\rm GeV}$
            & $4 \times 10^{5}\ {\rm GeV}$
            & $1 \times 10^{6}\ {\rm GeV}$\\
            10\ TeV (FO)
            & $3 \times 10^{7}\ {\rm GeV}$
            & $3 \times 10^{7}\ {\rm GeV}$
            & $2 \times 10^{7}\ {\rm GeV}$
            & $3 \times 10^{7}\ {\rm GeV}$\\
            10\ TeV (IT)
            & $8 \times 10^{8}\ {\rm GeV}$
            & $8 \times 10^{8}\ {\rm GeV}$
            & $6 \times 10^{8}\ {\rm GeV}$
            & $8 \times 10^{8}\ {\rm GeV}$\\
            10\ TeV (OS)
            & $4 \times 10^{9}\ {\rm GeV}$
            & $4 \times 10^{9}\ {\rm GeV}$
            & $3 \times 10^{9}\ {\rm GeV}$
            & $4 \times 10^{9}\ {\rm GeV}$\\
            \hline\hline
        \end{tabular}
        \caption{Upper bound on $T_{\rm R}$ for several values of the
        gravitino mass.  For the ${\rm D}$, ${\rm ^3He}$, and ${\rm
        ^6Li}$ abundances, here we use the observational constraints
        (\ref{lowd}), (\ref{He3/D}), (\ref{eq:obs6_upp}).  For ${\rm
        ^4He}$, we consider three cases (\ref{Y_FO}) (FO),
        (\ref{Y_IT}) (IT), (\ref{Y_OS}) (OS).}
        \label{table:T_R}
    \end{center}
\end{table}

Although Figs.\ \ref{fig:mtr1} $-$ \ref{fig:mtr4} look roughly the
same, the upper bound on the reheating temperature has model
dependences.  In particular, for a fixed value of the gravitino mass,
the upper bound depends on the MSSM parameters as one can see in the
figures.  To see this in more detail, in Table \ref{table:T_R}, we
show the upper bound on $T_{\rm R}$ for the cases listed in Table
\ref{table:mSUGRA}.  For the fixed value of the gravitino mass, the
upper bound on $T_{\rm R}$ may vary by the factor as large as $\sim
10$ when the gravitino mass is of the order of $1\ {\rm TeV}$.  When
the gravitino becomes heavier than $\sim 10\ {\rm TeV}$, however, the
upper bound becomes insensitive to the model parameters.  This is due
to the fact that, in this case, the branching ratios of the gravitino
are almost independent of the MSSM parameters.

Model-dependence of the upper bound on the reheating temperature is
mostly from the change of the lifetime and decay modes.  For the Case
1, we have chosen the mSUGRA parameters which give relatively light
superparticles (compared to other cases).  In the Case 2, masses of
all the superparticles are slightly increased compared to the Case 1.
Consequently, we can see some changes of the constraints on the
reheating temperature.

Compared to the Case 1, scalar masses are significantly increased in
the Case 3 with the gaugino masses being unchanged.  In this case,
gravitino is likely to decay into the gauginos, in particular, into
the gluino when kinematically allowed. (See Fig.\ \ref{fig:br}.)  We
found that the gluon-gluino final state produces more hadrons (in
particular, protons and neutrons) than the quark-squark final state.
Consequently, in the Case 3, upper bound on $T_{\rm R}$ becomes lower
than that for the Case 2.  We have also studied the case where masses
of all the squarks and sfermions are pushed to infinity by hand while
keeping the gaugino mass as low as $O(100\ {\rm GeV})$.  In this case,
the constraint on $T_{\rm R}$ is almost the same as that for the Case
3.  In addition, in the Case 4, masses of all the superparticles are
very large ($\sim$ a few TeV).  Then, lifetime of the gravitino
becomes relatively long, which makes the upper bound less stringent
for gravitinos with $m_{3/2}\sim$ a few TeV.

Although our main concern is to study the effects of the gravitino
decay on the BBN, it is also important to consider other constraints.
One of the important constraints is from the production of the LSP
from the decay of the gravitino.  Importantly, the LSP is produced
with the decay of the gravitino, and the present number density of the
LSP is given by the sum of two contributions; thermal relic, which is
calculated with the DarkSUSY package for each cases, and the
non-thermally produced LSP from the gravitino decay.  Since one LSP is
produced by the decay of one gravitino, the density parameter of the
LSP which has non-thermal origin is given by\footnote
{In fact, entropy production occurs when the gravitino decays, and
consequently, primordial LSPs are diluted by some amount.  For the
reheating temperature giving the constraint on the relic density of
the LSP, however, the effect of the dilution is negligible.}
\begin{eqnarray}
    \Delta \Omega_{\rm LSP} h^2 \simeq 0.054 \times
    \left( \frac{m_{\chi^0_1}}{100\ {\rm GeV}} \right)
    \left( \frac{T_{\rm R}}{10^{10}\ {\rm GeV}} \right),
\end{eqnarray}
where we have neglected the logarithmic corrections in Eq.\ 
(\ref{Ygrav}).  If we require, for example, that the total mass
density of the LSP be within the 95\ \% C.L. bound of the WMAP
constraint (i.e., $\Omega_{\rm LSP}^{\rm (thermal)}h^2+\Delta\Omega_{\rm
LSP}h^2<0.1287$ \cite{WMAP}), we also obtain upper bound on $T_{\rm
R}$, which is given by $3\times 10^{9}\ {\rm GeV}$, $1\times 10^{9}\ 
{\rm GeV}$, $4\times 10^{9}\ {\rm GeV}$, and $6\times 10^{8}\ {\rm
GeV}$, for the Cases 1, 2, 3, and 4, respectively.\footnote
{This bound is sensitive to the choice of the MSSM parameters since
the abundance of the thermally produced LSPs depends on the MSSM
parameter.  If we choose a parameter set which gives $\Omega_{\rm
LSP}^{\rm (thermal)}$ much smaller than the WMAP value, bound from the
production of the LSP may become much weaker.}
In our numerical analysis, we calculated the abundance
of the LSP taking account of the entropy production by the decay of
the gravitino; constraint from $\Delta\Omega_{\rm LSP}$ is also shown
in the figures.

Another constraint may be obtained from the distortion of the cosmic
microwave background (CMB).  An additional injection of the photon
into the thermal bath by the decaying particles is severely
constrained in order not to disturb the black-body shape of the CMB
spectrum \cite{Ellis:1990nb,Hu:1993gc}; numerically, $|\mu|< 9\times
10^{-5}$ for $\mu$-distortion and $|y|< 1.5\times 10^{-5}$ for
y-distortion are required \cite{Fixsen:1996nj}.  Using these
constraints, we obtain the upper bound on the total amount of the
injected energy $\Delta\rho_{\gamma}$; using the relation
$\Delta\rho_{\gamma}/s=E_{\rm vis}Y_{3/2}$,
\begin{eqnarray}
    \label{eq:mu-dist}
    \frac{\Delta \rho_{\gamma}}{s} <
    1.60 \times 10^{-13} {\rm GeV} \times
    \left(\frac{\tau_{3/2}}{10^{10}{\rm sec}} \right)^{-1/2} 
    \exp{\left[ (\tau_{\rm dC}/\tau_{3/2})^{5/4} \right]},
\end{eqnarray}
for $\mu$-distortion for $\tau_{\rm dC} \lesssim \tau_{3/2} \lesssim
2.5\times 10^{9}\ {\rm sec} \times (\Omega_{b}h^{2}/0.022)$
\cite{Hu:1993gc}, with $\tau_{\rm dC}$ being decoupling time of the
double Compton scattering:
\begin{eqnarray}
    \label{eq:tau_dC}
    \tau_{\rm dC} = 6.10 \times 10^{6} {\rm sec} \times  \left(
    \frac{T_{0}}{2.725~{\rm K}} \right)^{-12/5} \left(
    \frac{\Omega_b h^2}{0.022} \right)^{4/5} \left(
    \frac{1-Y_p/2}{0.88} \right)^{4/5},
\end{eqnarray}
and
\begin{eqnarray}
    \label{eq:y-dist}
    \frac{\Delta \rho_{\gamma}}{s}  <
    2.7 \times 
    10^{-13} {\rm GeV} \times 
    \left( \frac{\tau_{3/2}}{10^{10}\ {\rm sec}} \right)^{-1/2},
\end{eqnarray}
for $y$-distortion for $\tau_{3/2}\gtrsim 2.5\times 10^{9}\ {\rm sec}
\times (\Omega_{b}h^{2}/0.022)$ \cite{Ellis:1990nb}. Here $T_{0}$ is
the photon temperature at present.  Constraint from the distortion of
the CMB spectrum is also shown in the figures.

\subsection{Comment on the ${\rm ^7Li}$ abundance}
\label{sec:li7crisis}

So far, we have considered the constraints on the reheating
temperature, assuming that the prediction of the standard BBN agrees
with observations.  Although the standard BBN predicts the
light-element abundances which are more or less consistent with the
observational constraints, however, it has been pointed out that, if
we adopt the baryon-to-photon ratio suggested by the WMAP, standard
BBN predicts the ${\rm ^7Li}$ abundance slightly larger than the
observed value.  Indeed, if we do not include the additional
systematic error added in (\ref{eq:li7}), the standard BBN prediction
is found to be more than $2\sigma$ away from the center value. If we
take this discrepancy seriously, we need some explanation which may
include some effect of particle-physics model beyond the standard
model \cite{FenRajTak,Ichikawa:2004ju,Ichikawa:2004pb,Ellis:2005ii}.
Before closing this section, we comment on this issue.

If the net production of ${\rm ^7Li}$ can be somehow suppressed by the
decay of the long-lived particle (like the gravitino), the ${\rm
^7Li}$ discrepancy may be solved.  In the past, it was discussed that
the ${\rm ^7Li}$ abundance may be reduced by the photo-dissociation
process induced by the {\it radiative} decay of the long-lived
particle \cite{FenRajTak}.  The scenario with a long-lived particle
which decays only radiatively is, however, severely constrained by the
${\rm ^3He}$ constraint; in such a scenario, photo-dissociation of
background ${\rm ^4He}$ is also induced which overproduces ${\rm
^3He}$.  Thus, such a scenario does not work once the constraint on
the ${\rm ^3He}$ abundance is taken into account
\cite{Sigl:1995kk,Holtmann:1998gd,Kawasaki:2000qr}.

In order to solve the discrepancy, recently it was pointed out that
the suppression of the ${\rm ^7Li}$ abundance may be possible with
{\it hadronically} decaying long-lived particles.  In
\cite{Jedamzik:2004er}, it was discussed that, when the lifetime of
the long-lived particle is $\sim 10^3\ {\rm sec}$, abundance of ${\rm
^7Li}$ can become consistent with the observational constraint (with
no additional systematic error) without conflicting other constraints.
The reduction of ${\rm ^7Li}$ is mainly due to the dissociation of
${\rm ^7Be}$ (which decays into ${\rm ^7Li}$) by slow neutrons
produced in the hadronic shower.  (To be more exact, such slow
neutrons are supplied by the destruction of ${\rm ^4He}$, the
inter-conversion from protons, and so on.)

To study this issue, we have looked for the parameter region where the
${\rm ^7Li}$ abundance becomes consistent with the observational
constraint (\ref{Li7_Bonifacio}).\footnote
{Note in this case that the $\chi^{2}$ of $\log_{10}[(n_{\rm
^7Li}/n_{\rm H})]$ is calculated by Eq.~(\ref{eq:chi2_1}) (not
Eq.~(\ref{eq:chi2_2})), and 95 $\%$ C.L. corresponds to $\chi^{2} =
3.84$. In addition, because we fix the observational value of $n_{\rm
^6Li}/n_{\rm ^7Li}$, and we adopt the observational value of $n_{\rm
^7Li}/n_{\rm H}$ in (\ref{Li7_Bonifacio}), the observational
constraint on $n_{\rm ^6Li}/n_{\rm H}$ is also modified to be $(n_{\rm
^6Li} / n_{\rm H})^{\rm obs}<( 1.10^{+ 1.12}_{-0.56})\times 10^{-11}$
(2$\sigma$). }
Although dissociation of ${\rm ^7Be}$ by slow neutrons, which is the
most important process, is taken into account in our numerical
calculation, we could not include one of the non-thermal production
process of ${\rm ^7Li}$: $N+\alpha_{\rm BG}\rightarrow
N+\alpha+\pi's$, followed by $\alpha+\alpha_{\rm BG}\rightarrow {\rm
^7Li}+\cdots$.  This is because the experimental data for the energy
distribution of the final-state $\alpha$ is not available for the
first process.  Thus, the ${\rm ^7Li}$ abundance should be somehow
underestimated.  With a mild assumption that the kinetic energy of the
energetic $\alpha$ produced by the process $N+\alpha_{\rm
BG}\rightarrow N+\alpha+\pi's$ is $\sim 140\ {\rm MeV}$ in the center
of mass system independently of the energy of the beam nucleon
\cite{Epi140MeV}, however, we checked that the resultant ${\rm ^7Li}$
abundance is not significantly affected by this non-thermal production
process in the parameter region in which we will be interested.  Thus,
we believe that our calculation gives a reasonable estimate of the
${\rm ^7Li}$ abundance (even though the lower bound on the ${\rm
^7Li}$ abundance was not considered in deriving the upper bound on
$T_{\rm R}$).

\begin{figure}[t]
    \begin{center}
        \centerline{{\vbox{\epsfxsize=0.6\textwidth\epsfbox
        {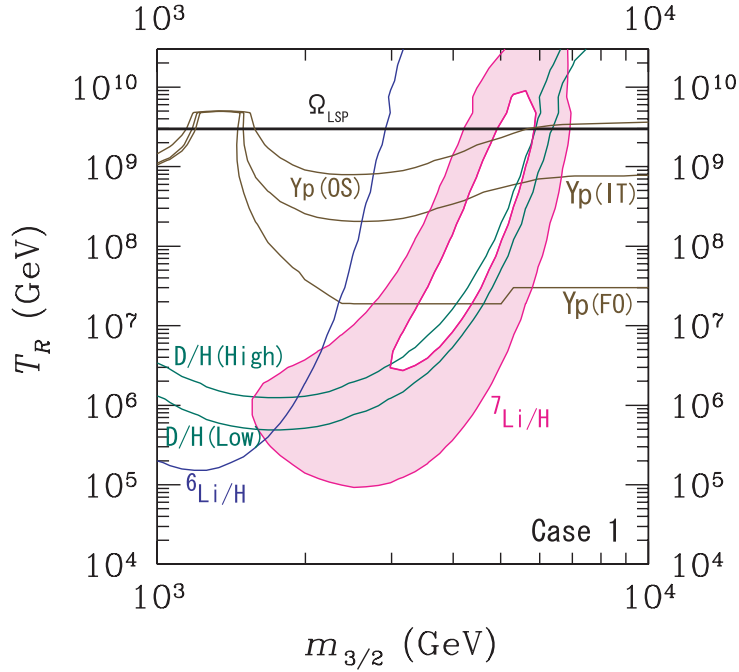}}}} 
        \caption{Parameter region which predicts the ${\rm ^7Li}$
        abundance consistent with (\ref{Li7_Bonifacio}); in the shaded
        region the ${\rm ^7Li}$ abundance becomes consistent with
        (\ref{Li7_Bonifacio}).  We have used the mSUGRA parameters for
        the Case 1.}
        \label{fig:Li7crisis}
    \end{center}
\end{figure}

In Fig.\ \ref{fig:Li7crisis}, we show the region where the ${\rm
^7Li}$ becomes consistent with the observational constraint
(\ref{Li7_Bonifacio}).  In this calculation, we have used the mSUGRA
parameters for the Case 1 to determine the MSSM parameters although
the result is insensitive to the choice of the mSUGRA parameters.  As
one can see, when $m_{3/2}\sim$ a few TeV and $T_{\rm R}\sim 10^{5-7}\ 
{\rm GeV}$ ($Y_{3/2}\sim 10^{-17} - 10^{-15}$ ), the ${\rm ^7Li}$
abundance becomes consistent with Eq.\ (\ref{Li7_Bonifacio}) without
conflicting the observational constraints for other light elements.
We have checked that this region is consistent with the parameter
region suggested in \cite{Jedamzik:2004er}.  We have also checked that
we can find a parameter region which predicts ${\rm ^7Li}$ abundance
consistent with (\ref{Li7_Ryan}).

\section{Conclusions and Discussion}
\setcounter{equation}{0}
\label{sec:conclusions}

In this paper, we have studied the effects of unstable gravitino on
the BBN in detail.  In particular, compared to the previous works, we
have performed the precise calculations of the decay rate and the
branching ratios of the gravitino.  For this purpose, we have first
fixed the masses and the mixing parameters of the MSSM particles, then
calculated the decay rates for all the relevant two and three body
decay processes of the gravitino.  Then, we calculate the spectrum of
the hadrons (in particular, $p$, $n$, and $\pi^\pm$).  With the hadron
spectrum as well as the visible energy emitted from the decay of the
gravitino, we calculate the light element abundances as functions of
the gravitino mass and the reheating temperature.  By comparing the
results of the theoretical calculation with the observational
constraints, we derived the upper bound on the reheating temperature
after the inflation.  

Although we have considered several difference mass spectrum of the
MSSM particles, the resultant constraints on the reheating temperature
behave qualitatively the same.  The detailed bound is, however,
sensitive to the mass spectrum of the superparticles and the upper
bounds on $T_{\rm R}$ for several cases are summarized in Table
\ref{table:T_R}.  When the gravitino mass is a few TeV, in particular,
the hadro-dissociation processes provide significant constraints.  Of
course, in some case, production of the hadrons are suppressed; in
particular, when the gravitino mass is close to the LSP mass, the only
possible two body decay process is
$\psi_\mu\rightarrow\gamma\chi^0_1$.  In this case, hadrons are
produced by the three body decay processes $\psi_\mu\rightarrow
q\bar{q}\chi^0_1$, which is suppressed compared to the two body decay
process.  Consequently, constraints become less stringent.

If the gravitino is the LSP, the gravitino becomes stable and the
cosmological constraints change drastically \cite{Moroi:1993mb,sWIMP}.
Detailed study of the case of the gravitino LSP will be given
elsewhere \cite{KohMorYot}

{\it Acknowledgment:} This work is supported in part by the 21st
century COE program, ``Exploring New Science by Bridging
Particle-Matter Hierarchy.''  The work of T.M. and K.K. is also
supported by the Grants-in Aid of the Ministry of Education, Science,
Sports, and Culture of Japan No.\ 15540247 (T.M.) and No.15-03605
(K.K.). K.K. is also supported by NSF grant AST 0307433.

\appendix

\section{Vertex Factors for the Gravitino Decay}
\setcounter{equation}{0}
\label{app:vertex}

In this appendix, we present the vertex factors for the decay
processes of the gravitino.  For the two-body decay processes with a
gauge boson in the final state, the decay rate can be calculated with
Eq.\ (\ref{Gamma_grav}) with Eq.\ (\ref{M2_gauge}) or
(\ref{M2_wzgauge}).  The vertex factors $C_L^{\rm (G)}$, $C_R^{\rm
(G)}$, $C_L^{\rm (H)}$, and $C_R^{\rm (H)}$ depends on the mixing
parameters.

For $\psi_\mu\rightarrow\gamma\chi_i^0$, the vertex factors are given
by 
\begin{eqnarray}
    \left[ C_L^{\rm (G)} \right]_{\psi_\mu\rightarrow\gamma\chi_i^0}
    = \left[ C_R^{{\rm (G)}*} \right]_{\psi_\mu\rightarrow\gamma\chi_i^0}
    =
    \frac{1}{g_Z} ( g_2  [ U_{\chi^0}^* ]_{i1}
    + g_1  [ U_{\chi^0}^* ]_{i2} ),
\end{eqnarray}
where $g_2$ and $g_1$ are gauge coupling constants for the $SU(2)_L$
and $U(1)_Y$ gauge group, respectively, while for the gluon-gluino
final state,
\begin{eqnarray}
    \left[ C_L^{\rm (G)} 
    \right]_{\psi_\mu\rightarrow g\tilde{g}}
    = \left[ C_R^{{\rm (G)} *} 
    \right]_{\psi_\mu\rightarrow g\tilde{g}}
    = 1.
\end{eqnarray}
In addition, for $\psi_\mu\rightarrow Z\chi_i^0$,
\begin{eqnarray}
    && 
    \left[ C_L^{\rm (G)} \right]_{\psi_\mu\rightarrow Z\chi_i^0}
    =\left[ C_R^{{\rm (G)}*} \right]_{\psi_\mu\rightarrow Z\chi_i^0}
    = \frac{1}{g_Z} ( - g_1  [ U_{\chi^0}^* ]_{i1}
    + g_2 [ U_{\chi^0}^* ]_{i2} ),
    \\ &&
    \left[ C_L^{\rm (H)} \right]_{\psi_\mu\rightarrow Z\chi_i^0}
    = -\left[ C_R^{{\rm (H)}*} \right]_{\psi_\mu\rightarrow Z\chi_i^0}
    = \frac{1}{\sqrt{2}} g_Z
    ( - v_1 [ U_{\chi^0}^* ]_{i3} + v_2 [ U_{\chi^0}^* ]_{i4} ),
\end{eqnarray}
and for $\psi_\mu\rightarrow W^\pm\chi_i^\mp$,
\begin{eqnarray}
    \left[ C_L^{\rm (G)} 
    \right]_{\psi_\mu\rightarrow W^\pm\chi_i^\mp}
    &=& [ U_{\chi^-}^* ]_{i1},
    \\
    \left[ C_R^{\rm (G)} 
    \right]_{\psi_\mu\rightarrow W^\pm\chi_i^\mp}
    &=& [ U_{\chi^+} ]_{i1},
    \\
    \left[ C_L^{\rm (H)} 
    \right]_{\psi_\mu\rightarrow W^\pm\chi_i^\mp}
    &=& - g_2 v_1 [ U_{\chi^-}^* ]_{i2},
    \\
    \left[ C_R^{\rm (H)} 
    \right]_{\psi_\mu\rightarrow W^\pm\chi_i^\mp}
    &=& g_2 v_2 [ U_{\chi^+} ]_{i2}.
\end{eqnarray}

For other processes, the decay rates can be calculated with Eq.\
(\ref{M2_chiral}).  The vertex factor for the neutral Higgs emission
processes ($\psi_\mu\rightarrow h\chi_i^0$, $\psi_\mu\rightarrow
H\chi_i^0$, and $\psi_\mu\rightarrow A\chi_i^0$) are given by
\begin{eqnarray}
    && \left[ C_L^{\rm (C)} 
    \right]_{\psi_\mu\rightarrow h\chi_i^0}
    = \left[ C_R^{{\rm (C)}*}
    \right]_{\psi_\mu\rightarrow h\chi_i^0}
    = 
    - \sin\alpha [ U_{\chi^0}^* ]_{i3} 
    + \cos\alpha [ U_{\chi^0}^* ]_{i4},
    \\
    && \left[ C_L^{\rm (C)} 
    \right]_{\psi_\mu\rightarrow H\chi_i^0}
    =  \left[ C_R^{{\rm (C)}*}
    \right]_{\psi_\mu\rightarrow H\chi_i^0}
    =  \cos\alpha [ U_{\chi^0}^* ]_{i3} 
    +  \sin\alpha [ U_{\chi^0}^* ]_{i4},
    \\
    && \left[ C_L^{\rm (C)} 
    \right]_{\psi_\mu\rightarrow A\chi_i^0}
    = - \left[ C_R^{{\rm (C)}*}
    \right]_{\psi_\mu\rightarrow A\chi_i^0}
    = \sin\beta [ U_{\chi^0}^* ]_{i3}
    + \cos\beta [ U_{\chi^0}^* ]_{i4} ,
\end{eqnarray}
while, for $\psi_\mu\rightarrow H^\pm\chi_i^\mp$,
\begin{eqnarray}
    \left[ C_L^{\rm (C)} 
    \right]_{\psi_\mu\rightarrow H^\pm\chi_i^\mp}
    &=& \sqrt{2} \sin\beta [ U_{\chi^-}^* ]_{i2},
    \\
    \left[ C_R^{\rm (C)} 
    \right]_{\psi_\mu\rightarrow H^\pm\chi_i^\mp}
    &=& \sqrt{2} \cos\beta [ U_{\chi^+} ]_{i2},
\end{eqnarray}
respectively.  For the rest of the processes (with a quark or a lepton
in the final state),
\begin{eqnarray}
    && \left[ C_L^{\rm (C)} 
    \right]_{\psi_\mu\rightarrow f \tilde{f}_i}
    = \sqrt{2} [U_{\tilde{f}}^{-1}]_{Li},
    \\
    && \left[ C_R^{\rm (C)} 
    \right]_{\psi_\mu\rightarrow f \tilde{f}_i}
    = \sqrt{2} [U_{\tilde{f}}^T]_{Ri}.
\end{eqnarray}

\section{Approximated Formula for Three-Body Process}
\setcounter{equation}{0}
\label{app:3body}

Even though there are several diagrams contributing to the three-body
decay process of the gravitino $\psi_\mu\rightarrow q\bar{q}\chi_1^0$,
photon-mediated diagram, Fig.\ \ref{fig:Feyn3body}(a), plays the most
important role when $m_{3/2}-m_{\chi_1^0}<m_Z$; indeed, in this case,
the decay rate $\Gamma(\psi_\mu\rightarrow q\bar{q}\chi_1^0)$ is well
approximated by the results only with Fig.\ \ref{fig:Feyn3body}(a).
Thus, although we have calculated all the relevant diagrams for the
three-body processes in our numerical study, we present the
approximated formula for the differential decay rate for the
three-body process, which is given by
\begin{eqnarray}
  \frac{d\Gamma(\psi_\mu\rightarrow q\bar{q}\chi_1^0)}
  {dm_{q\bar{q}}^2\,dm_{q\chi^0_1}^2} 
  \simeq
  \frac{d\Gamma(\psi_\mu\rightarrow\gamma^*\chi_1^0 
    \rightarrow q\bar{q}\chi_1^0)}
  {dm_{q\bar{q}}^2\,dm_{q\chi^0_1}^2} 
  =
  \frac{N_{\rm C}}{256\pi^3m_{3/2}^3M_*^2}\times
  \overline{\left| {\cal M} \right|^2},
\end{eqnarray}
where $N_{\rm C}=3$.  In addition, $m_{q\bar{q}}^2$ and
$m_{q\chi^0_1}^2$ are the invariant masses of the $q\bar{q}$ and
$q\chi^0_1$ systems, respectively, and are in the following range
\begin{alignat}{2}
  \big(2 m_q\big)^2 &\le
  m_{q\bar{q}}^2 \le
  \big( m_{3/2} - m_{\chi^0_1} \big)^2 ,\\
  \big( m_q + m_{\chi^0_1}\big)^2 &\le
  m_{q\chi^0_1}^2 \le
  \big(m_{3/2} - m_q \big)^2 ,\\
  m_{3/2}^2+m_{\chi^0_1}^2+2m_{q}^2 - \big( m_{3/2} - m_{q} \big)^2  &\le  
  m_{q\bar{q}}^2 + m_{q\chi^0_1}^2 \le
  m_{3/2}^2+m_{\chi^0_1}^2+2m_{q}^2 - \big( m_q + m_{\chi^0_1} \big)^2,
\end{alignat}
with $m_{q}$ being the mass of the final state quark.  The
photon-mediated three-body decay amplitude is given by
\begin{eqnarray}
  \overline{\left| {\cal M} \right|^2} &=&
  \frac{4}{3}\frac{e^2Q_{q}^2}{m_{q\bar{q}}^2}
  \left(C_L^{\rm{(G)}}C_L^{\rm{(G)}*}+C_R^{\rm{(G)}}C_R^{\rm{(G)}*}\right)
  \\
  && \bigg\{
  \left[ (kq')(kp)+(k' q')(k' p)
    - 2m_q^2 \left( (pq')-\frac{(pq)(qq')}{m_{q\bar{q}}^2} \right) \right] 
  \\ &&
  +\frac{(pq')}{m_{3/2}^2} \left[
    (pk)^2+(pk')^2 +2\frac{m_q^2(pq)^2}{m^2_{q\bar{q}}} \right]
  -m_{3/2}m_{\chi^0_1}\left(m_{q\bar{q}}^2 + 2m_q^2 \right) 
  \bigg\},
\end{eqnarray}
where $p$, $q'$, $k$, $k'$, $q$ are the momenta of $\psi_{\mu}$,
$\chi^0_1$, $q$, $\bar{q}$, and intermediate photon $\gamma^*$,
respectively.  Furthermore, $C^{\rm{(G)}}_L,C^{\rm{(G)}}_R$ are
defined in Appendix \ref{app:vertex}, and $eQ_q$ is the electric charge
of $q$.

As discussed in Section \ref{sec:decay}, when
$m_{3/2}-m_{\chi_1^0}>m_Z$, the process $\psi_\mu\rightarrow
q\bar{q}\chi_1^0$ is mostly mediated by the process with on-shell
$Z$-boson, and hence $\Gamma(\psi_\mu\rightarrow q\bar{q}\chi_1^0)$ is
well approximated by $\Gamma(\psi_\mu\rightarrow Z\chi_1^0)\times {\rm
Br}(Z\rightarrow q\bar{q})$.

\end{document}